\documentclass[onecolumn]{IEEEtran}

\usepackage{cite}

\ifCLASSINFOpdf
  \usepackage[pdftex]{graphicx}
\else
  \usepackage[dvips]{graphicx}
  \usepackage{epsfig}
\fi
\usepackage{amsfonts}
\usepackage{amsmath}
\usepackage{amssymb}
\usepackage{amscd}
\usepackage{mathrsfs}
\usepackage{color}
\usepackage{amsthm}
\usepackage{mathtools}
\allowdisplaybreaks
\usepackage{algorithmic}

\usepackage{array}

\usepackage{stfloats}
\usepackage{url}
\usepackage{eqparbox}
\usepackage[bookmarks=false]{hyperref}
\usepackage{fancyhdr}

\newtheorem{thrm}{Theorem}
\newtheorem{prop}{Proposition}
\newtheorem{lemma}{Lemma}

\newtheorem{remark}{Remark}

\begin{document}
%
\title{Learning to Detect an Odd Markov Arm}
%
%
%

\author{P. N. Karthik and
        Rajesh Sundaresan

        \thanks{The authors are with the Department of Electrical Communication Engineering at the Indian Institute of Science, Bangalore 560012, Karnataka, India.}

        \thanks{This work was supported by the Science and Engineering Research Board, Department of Science and Technology (grant no. EMR/2016/002503), and by the Robert Bosch Centre for Cyber Physical Systems at the Indian Institute of Science.}

        \thanks{This work was presented in part at the 2019 IEEE International Symposium on Information Theory (ISIT).}}
\maketitle

\begin{abstract}
A multi-armed bandit with finitely many arms is studied when each arm is a homogeneous Markov process on an underlying finite state space. The transition law of one of the arms, referred to as the odd arm, is different from the common transition law of all other arms. A learner, who has no knowledge of the above transition laws, has to devise a sequential test to identify the index of the odd arm as quickly as possible, subject to an upper bound on the probability of error. For this problem, we derive an asymptotic lower bound on the expected stopping time of any sequential test of the learner, where the asymptotics is as the probability of error vanishes. Furthermore, we propose a sequential test, and show that the asymptotic behaviour of its expected stopping time comes arbitrarily close to that of the lower bound. Prior works deal with independent and identically distributed arms, whereas our work deals with Markov arms.
Our analysis of the rested Markov setting is a key first step in understanding the difficult case of restless Markov setting, which is still open.

\end{abstract}


%
\IEEEpeerreviewmaketitle

\section{Introduction}\label{sec:introduction}
{\color{black} We study a multi-armed bandit problem with finitely many arms in which each arm is identified with a time homogeneous, irreducible and aperiodic discrete time Markov process on a finite state space. We assume that the state space is common to all the arms, and that the Markov process of any given arm is independent of the Markov process of every other arm. The state evolution on one of the arms is governed by a probability transition matrix $P_1$, while those on each of the other arms is governed by a probability transition matrix $P_2$, where $P_2\neq P_1$, hence making one of the arms anomalous (hereinafter referred to as the odd arm). A learner seeks to identify the index of the odd arm as quickly as possible, subject to an upper bound on the probability of error. We assume that the learner knows neither $P_1$ nor $P_2$, but knows that one of the arms is anomalous. We further assume that the learner can only devise sequential arm selection schemes in which, at each time, he may choose any one of the arms and observe the current state of the chosen arm. During this time, the unobserved arms do not undergo state transitions and remain frozen at their last observed states. We refer to this as the \emph{rested} arms setting, borrowing the terminology from Gittins \cite{gittins1979bandit}. Thus, our problem is one of odd arm identification in a multi-armed bandit setting with rested Markovian arms.
}

\subsection{\textcolor{black}{Prior Works That Deal With Rested and Markov Arms}}
 \textcolor{black}{One of the earliest works to consider the setting of rested and Markov arms is that of Gittins' \cite{gittins1979bandit} in which it is assumed that each arm yields a random `reward' when selected, and that successive rewards from any given arm constitute a Markov process. In this reward setting, the central problem is one of maximising the infinite horizon average discounted reward. For this problem, Gittins proposed and demonstrated the optimality of a simple index-based policy that, at each time, involves constructing an index for every arm based on the knowledge of the transition laws of the arms and selecting an arm with the largest index.}

 \textcolor{black}{Agarwal et al. \cite{agrawal1989asymptotically} consider a similar setting as Gittins', where each arm yields Markov rewards. However, unlike in  \cite{gittins1979bandit}, the authors of \cite{agrawal1989asymptotically} do not assume the knowledge of the transition laws of the arms. Instead, they assume that the transition law of each arm is parameterised by an unknown parameter belonging to a known, finite, parameter space. Define `regret' as the difference between the infinite time horizon expected sum of rewards generated by any policy and that generated by a policy which knows the parameters of the arms. The goal of the authors of \cite{agrawal1989asymptotically} is to design policies whose regret, in the asymptotic limit as time $n\to\infty$, is $o(n^\alpha)$ for every $\alpha>0$. For this problem, the authors of \cite{agrawal1989asymptotically} provide a lower bound in which the long-term regret grows asymptotically as $\log n$ times a multiplicative constant that captures the hardness of the problem. Furthermore, they propose a policy and demonstrate that it achieves the lower bound in the limit as $n\to \infty$.}

 \textcolor{black}{While the aforementioned works deal with reward maximisation, and the associated regret minimisation in the unknown parameters setting, our problem is one of {\em optimal stopping}. Our motivation to study the setting of rested and Markov arms in the context of odd arm identification comes from the fact that the lower bound in \cite{agrawal1989asymptotically}, although quantifying the asymptotic growth rate of regret, does not reflect the quickness of learning the underlying parameters of the arms. That is, the results in \cite{agrawal1989asymptotically} do not shed light on the minimum number of arm selections that are needed, on the average, in order to learn the parameters of the arms up to a desired level of accuracy. In this paper, we answer this question when one of the Markov arms is anomalous and the asymptotics is one of vanishing probability of error. In doing so, we treat the state of any selected arm as merely a Markov observation from the arm and not as a reward, since our objective is one of optimal stopping and not of regret minimisation. We note here that policies which are optimal in the context of the problem studied here may not necessarily be optimal in the context of regret minimisation, and vice-versa (we refer the reader to Bubeck et al. \cite{Bubeck2011} for a discussion on this). Finally, the unknown parameters of our problem are the transition laws of the odd arm and the non-odd arm Markov processes, and the index of the odd arm, thus making our parameter set a continuum, unlike the finite parameter set considered in \cite{agrawal1989asymptotically}.}

 \subsection{\textcolor{black}{Prior Works on Odd Arm Identification}}
 \textcolor{black}{The problem of odd arm identification is not new, and has been studied in the recent works of Vaidhiyan et al. \cite{vaidhiyan2018learning} for the case of independent and identically distributed (iid), indeed Poisson, observations from each arm, and of Prabhu et al. \cite{prabhu2017learning} for the case of iid observations belonging to a generic exponential family. {The works \cite{vaidhiyan2018learning} and \cite{prabhu2017learning}  can be embedded within the classical works of Chernoff \cite{Chernoff1959} and Albert \cite{albert1961sequential},  and provide a general framework for the analysis of lower bounds on expected number of samples required for identifying the index of the odd arm. In addition, they also provide explicit policies  that achieve these lower bounds in the asymptotic regime as error probability vanishes. We refer the reader to also \cite{garivier2016optimal, kaufmann2016complexity,hemo2016asymptotically,nitinawarat2017,naghshvar2010active,naghshvar2010information,naghshvar2011performance,naghshvar2013active} for other related works on iid observations.} While the aforementioned works deal with iid arms, the novelty in this paper is that we consider Markov arms. To the best of our knowledge, we believe that this work is the first to consider Markov arms in the context of odd arm identification.}

{\color{black}
\subsection{Our Motivation to Study the Rested Odd Markov Arm Problem}

  Vaidhiyan et al. \cite{Vaidhiyan2017} modeled visual search for locating an oddball image in a sea of distracter images, as quickly as possible, as an odd arm identification problem with Poisson observations. The Poisson observations stemmed from the Poisson point process model for the neuron firings when the learner focuses on a particular image, the analogue of pulling an arm. They showed that dissimilarity in neural responses to the oddball and the distracter images predicted the time taken by human subjects in detecting the location of the oddball image. The analysis was extended to the case when the parameters of the process were unknown, but had to be learnt during search, in Vaidhiyan et al. \cite{vaidhiyan2018learning}. The oddball and distracter images in the experiments analysed in Vaidhiyan et al. \cite{Vaidhiyan2017, vaidhiyan2018learning} and in Sripati and Olson \cite{sripati2010global} were static images. Similar experiments, but with dynamic drifting-dots images, as in Krueger et al. \cite{krueger2017evidence}, were conducted by Vaidhiyan et al. to see how evidence is accumulated in slow perceptual decision making. In these experiments, the dots executed Brownian motions with a fixed drift at each location. Moreover, the drifts were identical in the distracter locations and were different from the drift in the oddball location. Subjects had to identify the oddball location as quickly as possible. A proper analysis of this visual search, along the lines of \cite{Vaidhiyan2017} and \cite{vaidhiyan2018learning}, requires an understanding of the so-called \emph{restless} odd Markov arm problem where the unobserved arms continue to undergo state evolution. Indeed, in the aforementioned drifting-dots experiment, the state (positions of the dots) will have changed when the subject returns to observe a particular location after a decision to look at another location.

  There are other applications that can be modeled as the restless odd Markov arm problem, e.g., dynamic spectrum access in cognitive radio networks \cite{zhao2007survey}, single transmission line outages in power grids \cite{kumar2019sequential} but with limited observations, etc.

  The restless setting presents many analytical difficulties. As a key first step towards an understanding of the restless setting, our goal in this paper is to provide an analysis of the more tractable rested Markov arms setting. The rested case has its own interesting features. For example, as we shall see later in the paper, the asymptotically optimal arm selection strategy does not explicitly depend on the last observed states of the arms. This, at first glance, is surprising.

  Finally, a recent and independent work of Moulos \cite{moulos2019optimal} studies a closely related problem of \emph{best arm identification} in rested Markov multi-armed bandits, where the goal is to identify the arm with the largest stationary mean. The results presented in \cite{moulos2019optimal} are in terms of an asymptotic and a non-asymptotic lower bound, where the asymptotics is as the probability of error vanishes, and a policy for best arm identification whose asymptotic upper bound is four times larger than the asymptotic lower bound. In this paper, we present the first known asymptotic lower bound for the problem of odd arm identification, and an asymptotically optimal policy that meets the lower bound. This is in contrast with the gap between the upper bound and the lower bound in \cite{moulos2019optimal} for the best arm identification. We anticipate that a policy similar to ours should close the gap between the upper and lower bounds in \cite{moulos2019optimal}.
}

\subsection{Contributions}
Below, we highlight the key contributions of our work. Further, we mention the similarities and differences of our work with the aforementioned ones, and also bring out the challenges that we need to overcome in the analysis for the Markov setting.
\begin{enumerate}
    \item \textcolor{black}{In Section \ref{sec:lower_bound}}, we derive an asymptotic lower bound on the expected number of arm selections required by any policy that the learner may use to identify the index of the odd arm. Here, the asymptotics is as the error probability vanishes. Similar to the lower bounds appearing in \cite{agrawal1989asymptotically}-\cite{prabhu2017learning}, our lower bound has a problem-instance (or arms configuration) dependent constant that quantifies the effort required by any policy to identify the true index of the odd arm by guarding itself against identifying the nearest, incorrect alternative. \textcolor{black}{This constant is a function of the transition probability matrices of the odd arm and the non-odd arms.}
    \item We characterise the growth rate of the expected number of arm selections required by any policy as a function of the maximum acceptable error probability, and show that in the regime of vanishingly small error probabilitys, this growth rate is logarithmic in the inverse of the error probability. The analysis of the lower bounds in \cite{vaidhiyan2018learning} and \cite{prabhu2017learning} uses the familiar data processing inequality presented in the work of Kaufmann et al. \cite{kaufmann2016complexity} that is based on Wald's identity \cite{wald1944cumulative} for iid processes. However, the Markov setting in our problem does not permit the use of Wald's identity. Therefore, we derive results for our Markov setting generalising those appearing in \cite{kaufmann2016complexity}, and subsequently use these generalisations to arrive at the lower bound. See Section \ref{sec:lower_bound} for the details.
    \item \textcolor{black}{In the analysis of the lower bound, we bring out the key idea that any two successive selections of an arm result in the learner observing a transition from the state corresponding to the arm's first selection to the state corresponding to the arm's second selection. As a consequence, for each state in the state space, the empirical proportion of times an arm occupies the state prior to a transition is equal, in the long run, to the empirical proportion of times the arm occupies the state after a transition. We then replace these common proportions by the probability of the arm occupying this state under its stationary distribution. Such a replacement by stationary probabilities is possible mainly due to the rested nature of the arms, and may not be possible in more general settings such as when the arms are restless.}
    \item \textcolor{black}{In Section \ref{sec:achievability}}, we propose a sequential arm selection scheme that takes as inputs two parameters, one of which may be chosen appropriately to meet the acceptable error probability, while the other may be tuned to ensure that the performance of our scheme comes arbitrarily close to the lower bound, thus making our scheme near-optimal.

    We now contrast the near-optimality of our scheme with the near-optimality of the scheme proposed by Vaidhiyan et al. in \cite{vaidhiyan2018learning}, and highlight a key simplification that our scheme entails. The scheme of Vaidhiyan et al. is built around the important fact that each arm is sampled at a non-trivial, strictly positive and optimal rate that is bounded away from zero, as given by the lower bound, thereby allowing for exploration of the arms in an optimal manner. This stemmed from their specific Poisson observations. However, the lower bound presented in Section \ref{sec:lower_bound} may not have this property
in the context of Markov observations.
    Therefore, recognising the requirement of sampling the arms at a non-trivial rate for good performance of our scheme, in this paper, we use the idea of ``forced exploration'' proposed by Albert in \cite{albert1961sequential}. In particular, we propose a simplified way of sampling the arms by considering a mixture of uniform sampling and the optimal sampling given by the lower bound in Section \ref{sec:lower_bound}. We do this by introducing an appropriately tuneable parameter that controls the probability of switching between uniform sampling and optimal sampling, the latter being given by the lower bound. While this ensures that our policy samples each arm with a strictly positive probability at each step, it also gives us the flexibility to select an appropriate value for this parameter so that the upper bound on the performance of our scheme may be made arbitrarily close to our lower bound. We refer the reader to Section \ref{sec:achievability} for the details.

\end{enumerate}

\subsection{Organisation}
The rest of the paper is organised as follows. In Section \ref{sec:notations_and_preliminaries}, we set up some of the basic notations that will be used throughout the paper. In Section \ref{sec:lower_bound}, we present a lower bound on the performance of any policy. In Section \ref{sec:achievability}, we present a sequential arm selection policy and demonstrate its near optimality. We present the main result of this paper in Section \ref{sec:main_result}, combining the results of Sections \ref{sec:lower_bound} and \ref{sec:achievability}. In Section \ref{sec:simulation_results}, we provide some simulation to support the theoretical development, and provide concluding remarks in Section \ref{sec:conclusions}. We present the proofs of the main results in Section \ref{sec:proofs_of_main_results}.

\section{{\color{black}Notations and Preliminaries}} \label{sec:notations_and_preliminaries}
In this section, we set up the notations that will be used throughout the rest of this paper. Let $K\geq 3$ denote the number of arms, and let $\mathcal{A}=\{1,2,\ldots,K\}$ denote the set of arms. We associate with each arm an irreducible, aperiodic, time homogeneous discrete-time Markov process on a finite state space $\mathcal{S}$, where the Markov process of each arm is independent of the Markov processes of the other arms. We denote by $|\mathcal{S}|$ the cardinality of $\mathcal{S}$. Without loss of generality, we take $\mathcal{S}=\{1,2,\ldots,|\mathcal{S}|\}$. Hereinafter, we use the phrase `Markov process of arm $a$' to refer to the Markov process associated with arm $a\in \mathcal{A}$.

At each discrete time instant, one out of the $K$ arms is selected and its state is observed. We let $A_n$ denote the arm selected at time $n$, and let $\bar{X}_n$ denote the state of arm $A_n$, where $n\in\{0,1,2,\ldots\}$. We treat $A_0$ as the zeroth arm selection and $\bar{X}_0$ as the zeroth observation. Selection of an arm at time $n$ is based on the history $(\bar{X}^{n-1},A^{n-1})$ of past observations and arms selected; here, $\bar{X}^k$ (resp. $A^k$) is a shorthand notation for the sequence $\bar{X}_0,\ldots,\bar{X}_k$ (resp. $A_0,\ldots,A_k$). We shall refer to such a sequence of arm selections and observations as a policy, which we generically denote by $\pi$. For each $a\in\mathcal{A}$, we denote the Markov process of arm $a$ by the collection $(X_k^a)_{k\geq 0}$ of random variables. Further, we denote by $N_a(n)$ the number of times arm $a$ is selected by a policy up to (and including) time $n$, i.e.,
\begin{equation}
N_a(n)=\sum\limits_{t=0}^{n}1_{\{A_t=a\}}.\label{eq:N_a(n)}	
\end{equation}
 Then, for each $n\geq 0$, we have the observation
\begin{equation}
	\bar{X}_n=X_{N_{A_n}(n)-1}^{A_n}.\label{eq:obs_at_time_n}
\end{equation}

We consider a scenario in which the Markov process of one of the arms (hereinafter referred to as the odd arm) follows a probability transition matrix $P_1=(P_1(j|i))_{i,j\in\mathcal{S}}$, while those of rest of the arms follow a probability transition matrix $P_2=(P_2(j|i))_{i,j\in\mathcal{S}}$, where $P_2\neq P_1$; here, $P(j|i)$ denotes the entry in the $i${th} row and $j$th column of the matrix $P$. Further, we let $\mu_1$ and $\mu_2$ denote the unique stationary distributions of $P_1$ and $P_2$ respectively. We denote by $\nu$ the common distribution for the initial state of each Markov process. In other words, for arm $a\in \mathcal{A}$, we have $X_0^a\sim \nu$, and this is the same distribution for all arms. We operate in a setting where the probability transition matrices and their associated stationary distributions are unknown to the learner.

For each $a\in\mathcal{A}$ and state $i\in \mathcal{S}$, we denote by $N_a(n,i)$ the number of times up to (and including) time $n$ the Markov process of arm $a$ is observed to \textcolor{black}{occupy state $i$ prior to a transition}, i.e.,
\begin{equation}
	N_a(n,i)=\sum\limits_{m=1}^{N_a(n)-1} 1_{\{X_{m-1}^a=i\}}.\label{eq:N_a(n,i)}
\end{equation}
Similarly, for each $i,j\in \mathcal{S}$, we denote by $N_a(n,i,j)$ the number of times up to (and including) time $n$ the Markov process of arm $a$ is observed to \textcolor{black}{make a transition from state $i$ to state $j$}, i.e.,
\begin{equation}
	N_a(n,i,j)=\sum\limits_{m=1}^{N_a(n)-1} 1_{\{X_{m-1}^a=i,\,X_m^a=j\}}.\label{eq:N_a(n,i,j)}
\end{equation}
Clearly, then, the following hold:
\begin{subequations}
\begin{enumerate}
	\item For each $a\in \mathcal{A}$ and $i\in\mathcal{S}$,
	\begin{equation}
		\sum\limits_{j\in\mathcal{S}}N_a(n,i,j)=N_a(n,i).\label{eq:sum_N_a(n,i,j)}
	\end{equation}
	\item For each $a\in \mathcal{A}$,
	\begin{equation}
		\sum\limits_{i\in\mathcal{S}}N_a(n,i)=N_a(n)-1.\label{eq:N_a_summed_over_i}
	\end{equation}
	\item For each $n$,
	\begin{equation}
		\sum\limits_{a\in\mathcal{A}}N_a(n)=n+1.\label{eq:sum_N_a}
	\end{equation}
\end{enumerate}
\end{subequations}
We note here that the upper index of the summation in \eqref{eq:N_a(n,i)} is $N_a(n)-1$, and not $N_a(n)$, since the last observed transition on arm $a$ would be from the state $X_{N_a(n)-2}^a$ to the state $X_{N_a(n)-1}^a$. This is further reflected by the summation in \eqref{eq:N_a_summed_over_i}.

Fix probability transition matrices $P_1$ and $P_2$, where $P_2\neq P_1$, and let $H_{h}$ denote the hypothesis that $h$ is the index of the odd arm. The probability transition matrix of arm $h$ is $P_1$; all other arms have $P_2$. We refer to the triplet $C=(h,P_1,P_2)$ as a configuration. Our problem is one of detecting the true hypothesis among all possible configurations given by $$\mathcal{C}=\{C=(h,P_1,P_2):h\in\mathcal{A},\,\textcolor{black}{P_1\text{ and }P_2\text{ are transition probability matrices on }\mathcal{S}},\, \,P_2\neq P_1\}$$
when $P_1$ and $P_2$ are unknown. Let $C=(h,P_1,P_2)$ denote the underlying configuration of the arms. For each $a\in\mathcal{A}$, we denote by $(Z_h^a(n))_{n\geq 0}$ the log-likelihood process of arm $a$ under configuration $C$, with $h$ being the true index of the odd arm. Using the notations introduced above, we may then express $Z_h^a(n)$ as
\begin{equation}
	Z_h^a(n)=
\begin{cases}
	0,& N_a(n)=0,\\
	\log \nu(X_0^a),& N_a(n)=1,\\
	\log\nu(X_0^a)+\sum\limits_{m=1}^{N_a(n)-1}\log P_h^a(X_m^a|X_{m-1}^a),& N_a(n)\geq 2,
\end{cases}
\label{eq:Z_h^a(n)}
\end{equation}
where $P_h^a(j|i)$ denotes the conditional probability under hypothesis $H_h$ of observing state $j$ on arm $a$ given that state $i$ was observed on arm $a$ at the previous sampling instant, and is given by
\begin{equation}
	P_h^a(j|i)=\begin{cases}
		P_1(j|i),&a=h,\\
		P_2(j|i),&a\neq h.
	\end{cases}
	\label{eq:P_h^a(j|i)}
\end{equation}
Then, since the Markov processes of all the arms are independent of one another, for a given sequence $(A^n,\bar{X}^n)$ of arm selections and observations under a policy $\pi$ and a configuration $C=(h,P_1,P_2)$, denoting by $(Z_h(n))_{n\geq 0}$ the log-likelihood process under hypothesis $H_h$ of all arm selections and observations up to time $n$, we have
\begin{equation}
Z_h(n)=\sum\limits_{a=1}^{K}Z_h^a(n),\label{eq:log_likelihood_under_hyp_h}
\end{equation}
where $Z_h^a(n)$ is as given in \eqref{eq:Z_h^a(n)}.
On similar lines, for any two configurations $C=(h,P_1,P_2)$ and $C'=(h',P_1',P_2')$, where $P_2'\neq P_1'$ and $h'\neq h$, for each $a\in\mathcal{A}$, we define the log-likelihood process $(Z_{hh'}^a(n))_{n\geq 0}$ of configuration $C$ with respect to configuration $C'$ for arm $a$ as
\begingroup\allowdisplaybreaks\begin{align}
	Z_{hh'}^a(n)&=Z_h^a(n)-Z_{h'}^a(n)\nonumber\\
	&=\begin{cases}0,& N_a(n)=0,1,\\
	\sum\limits_{m=1}^{N_a(n)-1}\log \dfrac{P_{h}^{a}(X_m^a|X_{m-1}^a)}{P_{h'}^{a}(X_m^a|X_{m-1}^a)},& N_a(n)\geq 2.\label{eq:Z_{hh'}^a(n)}
\end{cases}
\end{align}\endgroup
We note that in the above equation, for $P_h^a$, we should use \eqref{eq:P_h^a(j|i)}, and for $P_{h'}^a$, we shall use, for all $a\in\mathcal{A}$ and $i,j\in\mathcal{S}$,
\begin{equation}
	P_{h'}^a(j|i)=\begin{cases}
		P_1'(j|i),&a=h',\\
		P_2'(j|i),&a\neq h'.
	\end{cases}
	\label{eq:P_{h'}^a(j|i)}
\end{equation}
Finally, we denote by $(Z_{hh'}(n))_{n\geq 0}$ the log-likelihood process of configuration $C$ with respect to $C'$ as
\begin{equation}
	Z_{hh'}(n)=\sum\limits_{a=1}^{K}Z_{hh'}^a(n),\label{eq:log_likelihood_under_hyp_h_and_h'}
\end{equation}
which includes all arm selections and observations.

The observation process $(\bar{X}_n)_{n\geq 0}$ and the arm selection process $(A_n)_{n\geq 0}$ are assumed to be defined on a common probability space $(\Omega,\mathcal{F},P)$. We define the filtration $(\mathcal{F}_n)_{n\geq 0}$ as
\begin{equation}
	\mathcal{F}_n = \sigma(A^n,\bar{X}^n),\quad n\geq 0.
	\label{eq:filtration}
\end{equation}
We use the convention that the zeroth arm selection $A_0$ is measurable with respect to the sigma algebra $\{\phi,\Omega\}$, whereas for all $n\geq 1$, the $n$th arm selection $A_n$ is $\mathcal{F}_{n-1}$-measurable.
For any stopping time $\tau$ with respect to the filtration in \eqref{eq:filtration}, we denote by $\mathcal{F}_{\tau}$ the $\sigma$-algebra
\begin{equation}
	\mathcal{F}_{\tau}=\{E\in\mathcal{F}:E\cap \{\tau=n\}\in\mathcal{F}_{n}\text{ for all }n\geq 0\}.\label{eq:stopping_time_sigma_alg}
\end{equation}

Our focus will be on policies $\pi$ that identify the index of the odd arm by sequentially sampling the arms, one at every time instant, and learning from the arms selected and observations obtained in the past. Specifically, at any given time, a policy $\pi$ prescribes one of the following alternatives:
\begin{enumerate}
\item Select an arm, based on the history of past observations and arms selected, according to a fixed distribution $\lambda$ independent of the underlying configuration of the arms, i.e., for each $n\geq 1$,
\begin{equation}
P(A_{n}=a|A^{n-1},\bar{X}^{n-1})=\lambda(a).	\label{eq:lambda(a)}
\end{equation}
\item Stop selecting arms, and declare the index $I(\pi)$ as the odd arm.	
\end{enumerate}
 Given a maximum acceptable error probability $\epsilon>0$, we denote by $\Pi(\epsilon)$ the family of all policies whose probability of error at stoppage for any underlying configuration of the arms is at most $\epsilon$. That is,
\begingroup\allowdisplaybreaks\begin{align}
	\Pi(\epsilon)=\bigg\lbrace\pi:P^\pi(I(\pi)\neq h|C)\leq \epsilon~\forall~ C=(h,P_1,P_2),\text{ where }h\in\mathcal{A}\text{ and }P_2\neq P_1\bigg\rbrace.\label{eq:Pi(epsilon)}
\end{align}\endgroup

For a policy $\pi$, we denote its stopping time by $\tau(\pi)$.
 Further, we write $E^\pi[\cdot|C]$ and $P^\pi(\cdot|C)$ to denote expectations and probabilities given that the underlying configuration of the arms is $C$. In this paper, we characterise the behaviour of $E^\pi[\tau(\pi)|C]$ for any policy $\pi\in\Pi(\epsilon)$, as $\epsilon$ approaches zero. We re-emphasise that $\pi$ cannot depend on the knowledge of $P_1$ or $P_2$, but could attempt to learn these along the way.

 \begin{remark}
Fix an odd arm index $h$, and consider the simpler case when $P_1$, $P_2$ are known, $P_2\neq P_1$. Let $\Pi(\epsilon|P_1,P_2)$ denote the set of all policies whose probability of error at stoppage is within $\epsilon$. From the definition of $\Pi(\epsilon)$ in \eqref{eq:Pi(epsilon)}, it follows that
\begin{equation}
\Pi(\epsilon)=\bigcap\limits_{P_1,P_2:P_2\neq P_1} \Pi(\epsilon|P_1,P_2). \label{eq:Pi(epsilon)_as_intersection}
\end{equation}
That is, policies in $\Pi(\epsilon)$ work for any $P_1, P_2$, with $P_2\neq P_1$. It is not a priori clear whether the set $\Pi(\epsilon)$ is nonempty. That it is nonempty for the case of iid observations was established in \cite{Chernoff1959}. In this paper, we show that $\Pi(\epsilon)$ is nonempty even for the setting of rested and Markov arms.
\end{remark}

\begin{remark}
	The distribution $\lambda$ appearing in \eqref{eq:lambda(a)} may, in general, be a function of time index $n$.
\end{remark}

In the next section, we provide a configuration dependent lower bound on $E^\pi[\tau(\pi)|C]$ for any policy $\pi\in\Pi(\epsilon)$. In Section \ref{sec:achievability}, we propose a sequential arm selection policy that achieves the lower bound asymptotically as the probability of error vanishes. We present the proofs in Section \ref{sec:proofs_of_main_results}.

\section{The Lower Bound}\label{sec:lower_bound}

 For any two transition probability matrices $P$ and $Q$ of dimension $|\mathcal{S}|\times |\mathcal{S}|$, and a probability distribution $\mu$ on $\mathcal{S}$, define $D(P||Q|\mu)$ as
\begin{equation}
 	D(P||Q|\mu)\coloneqq\sum\limits_{i\in\mathcal{S}}\mu(i)\sum\limits_{j\in\mathcal{S}}P(j|i)\log\frac{P(j|i)}{Q(j|i)},\label{eq:D(P_1||P|mu_1}
 \end{equation}
 with the convention $0\log 0=0\log\frac{0}{0}=0$. \textcolor{black}{The quantity in \eqref{eq:D(P_1||P|mu_1} is known as \emph{conditional informational divergence}, and the notation used above for representing the same is standard in the literature. See, for instance, Csisz\'{a}r and K\"{o}rner \cite[(2.4)]{csiszar2011information}.}

 The following proposition gives an asymptotic lower bound on the expected stopping time of any policy $\pi\in\Pi(\epsilon)$, as $\epsilon\downarrow 0$.
\begin{prop}\label{prop:lower_bound}
 	Let $C=(h,P_1,P_2)$ denote the underlying configuration of the arms. Then,
 	\begin{equation}
 		\lim\limits_{\epsilon\downarrow 0}\inf\limits_{\pi\in\Pi(\epsilon)}\frac{E^\pi[\tau(\pi)|C]}{\log ({1}/{\epsilon})}\geq \frac{1}{D^*(h,P_1,P_2)},\label{eq:lower_bound}
 	\end{equation}
 	where $D^*(h,P_1,P_2)$ is a configuration-dependent constant that is a function only of $P_1$ and $P_2$, and is given by
 	\begingroup\allowdisplaybreaks\begin{align}
 		D^*(h,P_1,P_2)
 		=\max\limits_{0\leq\lambda_1\leq 1}\left\lbrace\lambda_1\,D(P_1||P|\mu_1)+(1-\lambda_1)\frac{(K-2)}{(K-1)}D(P_2||P|\mu_2)\right\rbrace.\label{eq:D^*(h,P_1,P_2)_simpl_final}
 \end{align}\endgroup
 In \eqref{eq:D^*(h,P_1,P_2)_simpl_final}, $P$ is a probability transition matrix whose entry in the $i$th row and $j$th column is given by
 \begingroup\allowdisplaybreaks\begin{align}
 	P(j|i)=\frac{\lambda_1\mu_1(i)P_1(j|i)+(1-\lambda_1)\frac{(K-2)}{(K-1)}\mu_2(i)P_2(j|i)}{\lambda_1\mu_1(i)+(1-\lambda_1)\frac{(K-2)}{(K-1)}\mu_2(i)}.\label{eq:P(j|i)}
 \end{align}\endgroup
 \qed
 \end{prop}



{The proof of Proposition \ref{prop:lower_bound} broadly follows the outline of the proof of the lower bound in \cite{kaufmann2016complexity}, with necessary modifications for the setting of Markov rewards. We now outline some of the key steps in the proof. For an arbitrary choice of error probability $\epsilon>0$, we first show that for any policy $\pi\in\Pi(\epsilon)$,  the expected value of the total sum of log-likelihoods up to the stopping time $\tau(\pi)$ can be lower bounded by the binary relative entropy function
\begin{equation}
	d(\epsilon,1-\epsilon)\coloneqq\epsilon\log\frac{\epsilon}{1-\epsilon}+(1-\epsilon)\log\frac{1-\epsilon}{\epsilon}.\label{eq:d(epsilon,1-epsilon)}
\end{equation}}

{Next, we express the expected sum of log-likelihoods up to the stopping time $\tau(\pi)$ in terms of the expected value of the stopping time. It is in obtaining such an expression that works such as \cite{kaufmann2016complexity}, \cite{vaidhiyan2018learning} and \cite{prabhu2017learning} that are based on iid observations use Wald's identity, which greatly simplifies their analysis of the lower bound. Our setting of Markov rewards does not permit us to use Wald's identity. Therefore, we first obtain a generalisation of \cite[Lemma 18]{kaufmann2016complexity}, a change of measure based argument, to the setting of Markov rewards, and subsequently use this generalisation to obtain the desired relation.}

We then show that for any arm $a\in\mathcal{A}$, the long run frequency of observing the arm \textcolor{black}{occupying state $i\in \mathcal{S}$ prior to a transition} is equal to that of arm $a$ \textcolor{black}{occupying state $i$ after a transition}, and note that this common frequency is the stationary probability of observing the arm in state $i$. This explains the appearance of the unique stationary distributions $\mu_1$ and $\mu_2$ of the odd arm and the non-odd arms respectively in the expression \eqref{eq:D^*(h,P_1,P_2)_simpl_final}. We wish to emphasise that this step in the proof is possible due to the rested nature of the arms. The lower bound in the more general setting of ``restless'' arms in which the unobserved arms continue to undergo state transitions is still open.

{Finally, combining the above steps and using $d(\epsilon,1-\epsilon)/\log \frac{1}{\epsilon} \to 1$ as $\epsilon\downarrow 0$, we arrive at the lower bound in \eqref{eq:lower_bound}. The details may be found in Section \ref{appndx:proof_of_lower_bound}.}

\begin{remark}
	{The right-hand side of \eqref{eq:D^*(h,P_1,P_2)_simpl_final} is a function only of the probability transition matrices $P_1$ and $P_2$, and does not depend on the index $h$ of the odd arm. This is due to symmetry in the structure of arms, and we deduce that $D^*(h,P_1,P_2)$ does not depend on $h$. However, we include the index $h$ of the odd arm for the sake of consistency with the notation $C=(h,P_1,P_2)$ used to denote arm configurations. {\color{black} Further, it reminds us that $D^*$ may depend on all the parameters of the underlying configuration in more general composite hypothesis testing settings.}}
\end{remark}

 Going further, we let $\lambda^*\in[0,1]$ denote the value of $\lambda$ that achieves the maximum in \eqref{eq:D^*(h,P_1,P_2)_simpl_final}. We then define $\lambda_{opt}(h,P_1,P_2)=(\lambda_{opt}(h,P_1,P_2)(a))_{a\in\mathcal{A}}$ as the probability distribution on $\mathcal{A}$ given by
 \begin{equation}
 	\lambda_{opt}(h,P_1,P_2)(a)\coloneqq\begin{cases}
 		\lambda^*,&a=h,\\
 		\frac{1-\lambda^*}{K-1},&a\neq h.
 	\end{cases}\label{eq:lambda^*(h,P_1,P_2)}
 \end{equation}

 In the next section, we construct a policy that, at each time step, chooses arms with probabilities that match with those in \eqref{eq:lambda^*(h,P_1,P_2)} in the long run, in an attempt to reach the lower bound. While it is not a priori clear that this yields an asymptotically optimal policy, we show that this is indeed the case.



 \section{Achievability}\label{sec:achievability}
In this section, we propose a scheme that asymptotically achieves the lower bound of Section \ref{sec:lower_bound}, as the probability of error vanishes. Our policy is a modification of the policy proposed by Prabhu et al. \cite{prabhu2017learning} for the case of $K$ iid processes. We denote our policy by $\pi^\star(L,\delta)$, where $L\geq 1$ and $\delta\in(0,1)$ are the parameters of the policy.

Our policy is based on a modification of the classical generalised likelihood ratio (GLR) test in which we replace the maximum that appears in the numerator of the classical GLR statistic by an average computed with respect to a carefully constructed artificial prior over the space $\mathcal{P}(\mathcal{S})$ of all probability distributions on the state space $\mathcal{S}$. We describe this modified GLR statistic in the next section.

\subsection{The Modified GLR Statistic}
We revisit \eqref{eq:log_likelihood_under_hyp_h}, and suppose that each arm is selected once in the first $K$ time slots. Note that this does not affect the asymptotic performance. Then, under configuration $C=(h,P_1,P_2)$, the log-likelihood process $Z_h(n)$ may be expressed for any $n\geq K$ as
\begingroup\allowdisplaybreaks\begin{align}
	Z_h(n)=\sum\limits_{a=1}^{K}\log\nu(X_0^a)+\sum\limits_{i,j\in\mathcal{S}}N_h(n,i,j)\log P_1(j|i)
	+\sum\limits_{i,j\in\mathcal{S}}\left(\sum\limits_{a\neq h}N_a(n,i,j)\right)\log P_2(j|i),\label{eq:Z_{hh'}(N)_alt_expr}
\end{align}\endgroup
from which the likelihood process under $C$, denoted by $f(A^n,\bar{X}^n|C)$, may be written as
\begingroup\allowdisplaybreaks\begin{align}
	f(A^n,\bar{X}^n|C)=\prod\limits_{a=1}^K \nu(X_0^a)\prod\limits_{i,j\in\mathcal{S}}(P_1(j|i))^{N_h(n,i,j)}
	\cdot \prod\limits_{i,j\in\mathcal{S}}(P_2(j|i))^{\sum\limits_{a\neq h}N_a(n,i,j)}.\label{eq:likelihod_under_hyp_h}
\end{align}\endgroup

We now introduce an artificial prior on the space of all transition probability matrices for the state space $\mathcal{S}$. \textcolor{black}{Our choice of the prior is motivated by the requirement of having an appropriate conjugate prior for the likelihood in \eqref{eq:likelihod_under_hyp_h}. We therefore construct the Dirichlet distribution-based prior, noting that it meets our requirement.} Let $\text{Dir}(1,\ldots,1)$ denote the Dirichlet distribution with $|\mathcal{S}|$ parameters $\alpha_1,\ldots,\alpha_{|\mathcal{S}|}$, where $\alpha_j=1$ for all $j\in\mathcal{S}$. Then, denoting by $\mathscr{P}(\mathcal{S})$ the space of all transition probability matrices of size $|\mathcal{S}|\times |\mathcal{S}|$, we specify a prior on $\mathscr{P}(\mathcal{S})$ using the above Dirichlet distribution as follows: for any $P=(P(j|i))_{i,j\in\mathcal{S}}\in \mathscr{P}(\mathcal{S})$, $P(\cdot|i)$ is chosen according to the above Dirichlet distribution, independently of $P(\cdot|j)$ for all $j\neq i$. Further, for any two matrices $P,Q\in\mathscr{P}(\mathcal{S})$, the rows of $P$ are independent of those of $Q$. Then, it follows that under this prior, the joint density at ($P_1$, $P_2$) for $P_1,P_2\in\mathscr{P}(\mathcal{S})$  is
\begingroup\allowdisplaybreaks\begin{align}
	\mathscr{D}(P_1,P_2)&\coloneqq\prod\limits_{i\in\mathcal{S}}\frac{\prod\limits_{j\in\mathcal{S}}(P_1(j|i))^{\alpha_j-1}}{B(1\ldots,1)}\prod\limits_{i\in\mathcal{S}}\frac{\prod\limits_{j\in\mathcal{S}}(P_2(j|i))^{\alpha_j-1}}{B(1\ldots,1)}\nonumber\\
	&=\frac{1}{B(1,\ldots,1)^{2|\mathcal{S}|}}\label{eq:Dirichlet_prior},
\end{align}\endgroup
where $B(1,\ldots,1)$ denotes the normalisation factor for the distribution $\text{Dir}(1,\ldots,1)$, and the second line above follows by substituting $\alpha_j=1$, $j\in\mathcal{S}$.

{\color{black} By a minor abuse of notation,} we denote by $f(A^n,\bar{X}^n|H_h)$ the average of the likelihood in \eqref{eq:likelihod_under_hyp_h} computed with respect to the prior in \eqref{eq:Dirichlet_prior}. From the property that the Dirichlet distribution is the appropriate conjugate prior for the observation process,
\begingroup\allowdisplaybreaks\begin{align}
	&f(A^n,\bar{X}^n|H_h)=\prod\limits_{a=1}^{K}\nu(X_0^a)\prod\limits_{i\in\mathcal{S}}\frac{B((N_h(n,i,j)+1)_{j\in\mathcal{S}})}{B(1,\ldots,1)}
	\prod\limits_{i\in\mathcal{S}}\frac{B\left(\left(\sum\limits_{a\neq h}N_a(n,i,j)+1\right)_{j\in\mathcal{S}}\right)}{B(1,\ldots,1)},\label{eq:f(Z^n,A^n|H_h)}
\end{align}\endgroup
where in the above expression, $B((N_h(n,i,j)+1)_{j\in\mathcal{S}})$ denotes the normalisation factor for a Dirichlet distribution with parameters $(N_h(n,i,j)+1)_{j\in\mathcal{S}}$. It can be shown that  $f(A^n,\bar{X}^n|H_h)$ is also the expected value of the likelihood in \eqref{eq:likelihod_under_hyp_h} computed with respect to the prior in \eqref{eq:Dirichlet_prior}, i.e.,
\begingroup\allowdisplaybreaks\begin{align}
	f(A^n,\bar{X}^n|H_h)
	=\prod\limits_{a=1}^{K}\nu(X_0^a)\prod\limits_{i\in\mathcal{S}}E\left[\prod\limits_{j\in\mathcal{S}}X_{ij}^{N_h(n,i,j)}\cdot Y_{ij}^{\sum\limits_{a\neq h}N_a(n,i,j)}\right]
\end{align}\endgroup
where in the above set of equations, the random vectors $(X_{ij})_{i,j\in\mathcal{S}}$ and $(Y_{ij})_{i,j\in\mathcal{S}}$ are independent with independent components, and jointly distributed according to \eqref{eq:Dirichlet_prior}, and the expectation is also with respect to this joint density.

Let $\hat{P}^n_{h,1}$ and $\hat{P}^n_{h,2}$ denote the maximum likelihood estimates of probability transition matrices $P_1$ and $P_2$ respectively, under hypothesis $H_h$. {Taking partial derivatives of the right-hand side \eqref{eq:likelihod_under_hyp_h} with respect to $P_1(j|i)$ and $P_2(j|i)$ for each $i,j\in\mathcal{S}$, and setting each of these derivatives to zero, we get
\begingroup\allowdisplaybreaks\begin{align}
	\hat{P}_{h,1}^n(j|i)
	=\frac{N_h(n,i,j)}{N_h(n,i)},\quad
	\hat{P}_{h,2}^n(j|i)
	=\frac{\sum\limits_{a\neq h}N_a(n,i,j)}{\sum\limits_{a\neq h}N_a(n,i)}.\label{eq:ML_est_under_hyp_h}
\end{align}\endgroup}


{Plugging the estimates in \eqref{eq:ML_est_under_hyp_h}  back into \eqref{eq:likelihod_under_hyp_h}, we get the maximum likelihood of all observations and actions under hypothesis $H_h$:}
\begingroup\allowdisplaybreaks\begin{align}
	\hat{f}(A^n,\bar{X}^n|H_h)&\coloneqq\max\limits_{C=(h,\cdot,\cdot)}f(A^n,\bar{X}^n|C)\nonumber\\
	&=\prod\limits_{a=1}^K \nu(X_0^a)\prod\limits_{i,j\in\mathcal{S}}\bigg\lbrace\left(\frac{N_h(n,i,j)}{N_h(n,i)}\right)^{N_h(n,i,j)}
	\left(\frac{\sum\limits_{a\neq h}N_a(n,i,j)}{\sum\limits_{a\neq h}N_a(n,i)}\right)^{\sum\limits_{a\neq h}N_a(n,i,j)}\bigg\rbrace.\label{eq:ml_likelihod_under_hyp_h}
\end{align}\endgroup

We now define our modified GLR statistic. Let $H_h$ and $H_{h'}$ be any two hypotheses, with $h'\neq h$. Let $\pi$ be a policy whose sequence of arm selections and observations up to (and including) time  $n$ is $(A^n,\bar{X}^n)$. Then, the modified GLR statistic of $H_h$ with respect to $H_{h'}$ up to time $n$ is denoted by $M_{hh'}(n)$, and is defined as
\begingroup\allowdisplaybreaks\begin{align}
	M_{hh'}(n)
	&=\log\frac{f(A^n,\bar{X}^n|H_h)}{\hat{f}(A^n,\bar{X}^n|H_{h'})}\nonumber\\
	&=T_1+T_2(n)+T_3(n)+T_4(n)+T_5(n),\label{eq:M_{hh'}(n)}
\end{align}\endgroup
where the terms appearing in \eqref{eq:M_{hh'}(n)} are as follows.
\begin{enumerate}
	\item The term $T_1$ is given by
	\begin{equation}
		T_1=2|\mathcal{S}|\log\left(\frac{1}{B(1,\ldots,1)}\right).\label{eq:t_1}
	\end{equation}
	\item The term $T_2(n)$ is given by
    \begin{equation}
    	T_2(n)=\sum\limits_{i\in\mathcal{S}}\log B((N_h(n,i,j)+1)_{j\in\mathcal{S}}).\label{eq:t_2(n)}
    \end{equation}
	\item The term $T_3(n)$ is given by
	\begin{equation}
		T_3(n)=\sum\limits_{i\in\mathcal{S}}\log B\left(\left(\sum\limits_{a\neq h}N_a(n,i,j)+1\right)_{j\in\mathcal{S}}\right).\label{eq:t_3(n)}
	\end{equation}
	\item The term $T_4(n)$ is given by
	\begin{equation}
		T_4(n)=-\sum\limits_{i,j\in\mathcal{S}}N_{h'}(n,i,j)\log\frac{N_{h'}(n,i,j)}{N_{h'}(n,i)}.\label{eq:t_4(n)}
	\end{equation}
	\item The term $T_5(n)$ is given by
	\begingroup\allowdisplaybreaks\begin{align}
		T_5(n)=-\sum\limits_{i,j\in\mathcal{S}}\sum\limits_{a\neq h'}N_{a}(n,i,j)\log\frac{\sum\limits_{a\neq h'}N_{a}(n,i,j)}{\sum\limits_{a\neq h'}N_{a}(n,i)}.\label{eq:t_5(n)}
	\end{align}\endgroup
\end{enumerate}

Note that $\nu$, the distribution of the initial state of any arm, is irrelevant since it appears in both \eqref{eq:f(Z^n,A^n|H_h)} and \eqref{eq:ml_likelihod_under_hyp_h}, and thus cancels out in writing \eqref{eq:M_{hh'}(n)}. Let us emphasise that our modified GLR statistic is one in which the maximum in the numerator of the usual GLR statistic is replaced by an average in \eqref{eq:f(Z^n,A^n|H_h)} computed with respect to the artificial prior over the space $\mathscr{P}(\mathcal{S})$ introduced in \eqref{eq:Dirichlet_prior}.

\begin{remark}
	We wish to mention here that the expression on the right-hand side of \eqref{eq:likelihod_under_hyp_h} for $f(A^n,\bar{X}^n|C)$ represents the likelihood of all observations up to (and including) time $n$ {``conditioned on''} the actions $A^n$ up to (and including) time $n$. In other words, a more precise expression for $f(A^n,\bar{X}^n|C)$ is as follows:
	\begingroup\allowdisplaybreaks\begin{align}
		f(A^n,\bar{X}^n|C)=\bigg[\prod\limits_{t=0}^n P_h(A_t|A^{t-1},\bar{X}^{t-1})\bigg]\prod\limits_{a=1}^K \nu(X_0^a)\prod\limits_{i,j\in\mathcal{S}}(P_1(j|i))^{N_h(n,i,j)}
	\cdot \prod\limits_{i,j\in\mathcal{S}}(P_2(j|i))^{\sum\limits_{a\neq h}N_a(n,i,j)},\label{eq:likelihod_under_hyp_h_precise}
	\end{align}\endgroup
	where $P_h(A_t|A^{t-1},\bar{X}^{t-1})$ represents the probability of selecting arm $A_t$ at time $t$ when the true hypothesis is $H_h$ (i.e., when $h$ is the index of the odd arm), with the convention that at time $t=0$, this term represents $P_h(A_0)$. Note that for any policy (see description in the paragraph containing
\eqref{eq:lambda(a)} and \eqref{eq:Pi(epsilon)}), this must be independent of the true hypothesis $H_h$, and
is thus the same for any two hypotheses $H_h$ and $H_{h'}$, where $h' \neq h$.
	
	
	As a consequence of this, the first term within square brackets on the right-hand side of \eqref{eq:likelihod_under_hyp_h_precise} appears in both the numerator and the denominator terms of the modified GLR statistic of \eqref{eq:M_{hh'}(n)}, and  thus cancels out. Hence, we omit writing this term in the expressions of \eqref{eq:likelihod_under_hyp_h}, \eqref{eq:f(Z^n,A^n|H_h)} and \eqref{eq:ml_likelihod_under_hyp_h}.
\end{remark}

\subsection{The Policy $\pi^\star(L,\delta)$}
With the above ingredients in place, we now describe our policy based on the modified GLR statistic of \eqref{eq:M_{hh'}(n)}. Let
\begin{equation}
	M_h(n)\coloneqq \min\limits_{h'\neq h}M_{hh'}(n)\label{eq:M_h(n)}
\end{equation}
denote the modified GLR of hypothesis $H_h$, $h\in\mathcal{A}$, with respect to its nearest alternative.
\vspace{0.2cm}\\
\emph{Policy }$\pi^\star(L,\delta)$:\\
Fix parameters $L\geq 1$ and $\delta\in(0,1)$. Let $(B_n)_{n\geq 1}$ be a sequence of iid Bernoulli($\delta)$ random variables such that $B_{n+1}$ is independent of the sequence $(A^n,\bar{X}^n)$ for all $n\in\{0,1,2,\ldots\}$.
We choose each of the $K$ arms once in the first $K$ time steps $n=0,\ldots,K-1$.  For each $n\geq K-1$, at time $n$, we follow the procedure described below:
\begin{enumerate}
	\item Let $h^{*}(n)=\arg\max\limits_{h\in\mathcal{A}}M_h(n)$ be the index with the largest modified GLR after $n$ time steps. We resolve ties uniformly at random.
	\item If $M_{h^*(n)}(n)<\log((K-1)L$, then we choose the next arm $A_{n+1}$ based on the sequence $(A^n,\bar{X}^n)$ of observations and arms selected until time $n$ as per the following rule:
	    \begin{enumerate}
            \item If $B_{n+1}=1$, then we choose an arm uniformly at random.
            \item If $B_{n+1}=0$, then we choose $A_{n+1}$ according to the distribution $\lambda_{opt}(h^*(n),\hat{P}^n_{h^*(n),1},\hat{P}^n_{h^*(n),2})$.
	    \end{enumerate}	
	\item If $M_{h^*(n)}(n)\geq\log((K-1)L)$, then we stop selecting arms and declare $h^*(n)$ as the true index of the odd arm.
\end{enumerate}

{In the above policy, $h^*(n)$ provides the best estimate  of the odd arm at time $n$. If the modified GLR statistic of arm $h^*(n)$ is sufficiently larger than that of its nearest incorrect alternative ($\geq \log((K-1)L)$), then this indicates that the policy is confident that $h^*(n)$ is the odd arm. At this stage, the policy stops taking further samples and declares $h^*(n)$ as the index of the odd arm. If not, the policy continues to obtain further samples.}

{We refer to the rule in item (2) above as \emph{forced exploration} with parameter $\delta$. A similar rule also appears in \cite{albert1961sequential}. Based on the description in items (2(a)) and (2(b)) above, it follows that for each $a\in\mathcal{A}$,}
\begingroup\allowdisplaybreaks\begin{align}
	P(A_{n+1}=a|A^n,\bar{X}^n)
	&=\frac{\delta}{K}+(1-\delta)\,\lambda_{opt}(h^*(n),\hat{P}^n_{h^*(n),1},\hat{P}^n_{h^*(n),2})(a)\nonumber\\
	&\geq \frac{\delta}{K}>0.\label{eq:P(A_{n+1}=a|sigma(A^n,X^n)}
\end{align}\endgroup
As we will see, the strictly positive lower bound in \eqref{eq:P(A_{n+1}=a|sigma(A^n,X^n)} will ensure that the policy selects each arm at a non-trivial frequency so as to allow for sufficient exploration of all arms. Also, we will show that the parameters $L$ and $\delta$ may be selected so that our policy achieves a desired target error probability, while also ensuring that the normalised expected stopping time of the policy is arbitrarily close to the lower bound in \eqref{eq:lower_bound}.

{
\begin{remark}
Evaluating the distribution $\lambda_{opt}(h^*(n),\hat{P}^n_{h^*(n),1},\hat{P}^n_{h^*(n),2})$ in step (2(a)) of the policy involves solving the maximisation problem in \eqref{eq:D^*(h,P_1,P_2)_simpl_final} with the probability transition matrices $P_1$ and $P_2$ replaced by their corresponding ML estimates $\hat{P}^n_{h^*(n),1}$ and $\hat{P}^n_{h^*(n),2}$ respectively at each time $n\geq K-1$ until stoppage. In the event when any of the rows of the estimated matrices has all its entries as zero, we substitute the corresponding zero row by a row with a single `1' in one of the $|\mathcal{S}|$ positions picked uniformly at random. Since the ML estimates converge to their respective true values as more observations are accumulated, we note that such a substitution operation (or any modification thereof that replaces the all-zero rows by an arbitrary probability vector) needs to be carried out only for finitely many time slots, and does not affect the asymptotic performance of the policy.
\end{remark}}

\subsection{Performance of $\pi^\star(L,\delta)$}
In this subsection, we show that the expected number of samples required by policy $\pi^\star(L,\delta)$ to identify the index of the odd arm can be made arbitrarily close to that in \eqref{eq:lower_bound} in the regime of vanishing error probabilities. We show that this can be achieved by choosing the parameters $L$ and $\delta$ carefully. We organise this subsection as follows:
\begin{enumerate}
	\item First, we show that when the true index of the odd arm is $h$, the modified GLR $M_h(n)$ of hypothesis $H_h$ with respect to its nearest alternative has a strictly positive drift under our policy. We then use this to show that our policy stops in finite time with probability $1$.
	\item For any fixed target error probability $\epsilon>0$, we show that  for an appropriate choice of the threshold parameter $L$, our policy belongs to the family $\Pi(\epsilon)$, i.e., its probability of error at stoppage is within $\epsilon$.
	\item We obtain an upper bound on the expected stopping time of our policy, and demonstrate that this upper bound may be made arbitrarily close to the lower bound in \eqref{eq:lower_bound} by choosing an appropriate value of $\delta\in (0,1)$.
\end{enumerate}

\subsubsection{Strictly Positive Drift of the Modified GLR Statistic}\label{subsubsec:positive_drift}
The main result on the strictly positive drift of the modified GLR statistic is as described in the following proposition.
\begin{prop}\label{prop:positive_drift_of_M_{hh'}(n)}
	Fix $L\geq 1$, $\delta\in(0,1)$, and consider a version of the policy $\pi^\star(L,\delta)$ that never stops. Let $C=(h,P_1,P_2)$ be the underlying configuration of the arms. Then, for all $h'\neq h$, under the non-stopping version of our policy, we have
	\begin{equation}
		\liminf\limits_{n\to\infty}\frac{M_{hh'}(n)}{n}>0.\label{eq:positive_drift_of_M_{hh'}(n)}
	\end{equation}
	\qed
\end{prop}

{The proof is based on the key idea that forced exploration with parameter $\delta\in(0,1)$ (see items (2(a)) and (2(b)) of policy $\pi^\star(L,\delta)$) results in sampling each arm with a strictly positive rate that grows linearly. It is in showing an analogue of Proposition \ref{prop:positive_drift_of_M_{hh'}(n)} for iid Poisson observations that the authors of \cite{vaidhiyan2018learning} use their result of \cite[Proposition 3]{vaidhiyan2018learning} on guaranteed exploration at a strictly positive rate. Since it is not clear if the analogue of \cite[Proposition 3]{vaidhiyan2018learning} holds in general, we use the idea in \cite{albert1961sequential} of forced exploration. We present the details in Section \ref{appndx:proof_of_strictly_positive_drift_of_M_{hh'}(n)}. We refer the reader to \cite{garivier2016optimal} on how to make do with forced exploration at a sublinear rate.}

As an immediate consequence of the above proposition, we have the following: suppose $C=(h,P_1,P_2)$ is the underlying configuration of the arms. Then, a.s.,
\begingroup\allowdisplaybreaks\begin{align}
	\liminf\limits_{n\to\infty}M_h(n)=\liminf\limits_{n\to\infty}\min\limits_{h'\neq h}M_{hh'}(n)>0.\label{eq:positive_drift_of_M_h(n)}
\end{align}\endgroup
The result in \eqref{eq:positive_drift_of_M_h(n)} has the following implication. For any $h'\neq h$, we have the following set of inequalities holding a.s.:
\begingroup\allowdisplaybreaks\begin{align}
	\limsup\limits_{n\to\infty}M_{h'}(n)&=\limsup\limits_{n\to\infty}\min\limits_{a\neq h'}M_{h'a}(n)\nonumber\\
	&\leq \limsup\limits_{n\to\infty}M_{h'h}(n)\nonumber\\
	&=\limsup\limits_{n\to\infty}-M_{hh'}(n)\nonumber\\
	&= -\liminf\limits_{n\to\infty}M_{hh'}(n)\nonumber\\
	&\leq -\liminf\limits_{n\to\infty}M_{h}(n)\nonumber\\
	&<0.\label{eq:limsup_M_{h'}(n)_less_than_0}
\end{align}\endgroup
From the above set of inequalities, it follows that under policy $\pi^\star(L,\delta)$,
 \begin{equation}
 	h^*(n)=\arg\max\limits_{h\in\mathcal{A}}M_h(n)=h\text{ a.s.}\label{eq:h^*(n)_equal_to_h_almost_surely}
 \end{equation}
for all sufficiently large values of $n$.

We note here that when $C=(h,P_1,P_2)$ is the underlying configuration of the arms, \eqref{eq:h^*(n)_equal_to_h_almost_surely} seems to suggest that  policy $\pi^\star(L,\delta)$ a.s. outputs $h$ as the true index of the odd arm at the time of stopping, thereby implying that it commits no error a.s. However, we wish to remark that this is not true, and recognise the possibility of the event that $h^*(n)=h'\neq h$ and $M_{h^*(n)}(n)\geq \log((K-1)L)$  for some $n$, in which case the policy stops at time $n$ and outputs $h'$ as the index of the odd arm, thereby making error. While we shall soon demonstrate that the probability of occurrence of such an error event under our policy is small, we leverage the implication of \eqref{eq:h^*(n)_equal_to_h_almost_surely} to define a version of our policy that, under the underlying configuration $C=(h,P_1,P_2)$, waits until the event $M_h(n)\geq \log((K-1)L)$ occurs, at which point it stops and declares $h$ as the index of the odd arm. We denote this version by $\pi^\star_h(L,\delta)$. Thus, $\pi^\star_h(L,\delta)$ stops only at declaration $h$.

It then follows that the stopping times of policies $\pi^\star(L,\delta)$ and $\pi^\star_h(L,\delta)$ are a.s. related as $\tau(\pi^\star_h(L,\delta))\geq \tau(\pi^\star(L,\delta))$, as a consequence of which we have the following set of inequalities holding a.s.:
\begingroup\allowdisplaybreaks\begin{align}
	\tau(\pi^\star(L,\delta))\leq \tau(\pi^\star_h(L,\delta))
	&=\inf\{n\geq 1:M_h(n)\geq \log((K-1)L)\}\nonumber\\
	&\leq \inf\bigg\lbrace n\geq 1:M_{hh'}(n')\geq \log((K-1)L)\text{ for all }n'\geq n\text{ and for all }h'\neq h\bigg\rbrace\nonumber\\
	&<\infty,\label{eq:stopping_time_finite_almost_surely}
\end{align}\endgroup
where the last line follows as a consequence of Proposition \ref{prop:positive_drift_of_M_{hh'}(n)}. This establishes that a.s. policy $\pi^\star(L,\delta)$ stops in finite time.

\subsubsection{Probability of Error of Policy $\pi^\star(L,\delta)$}
We now show that for policy $\pi^\star(L,\delta)$, the threshold parameter $L$ may be chosen to achieve any desired target error probability. This is formalised in the proposition below.
\begin{prop}\label{prop:pi_{LRMB}(L,delta)_belongs_to_Pi(epsilon)}
	Fix $\epsilon>0$. Then, for $L=1/\epsilon$, we have $\pi^\star(L,\delta)\in\Pi(\epsilon)$ for all $\delta\in(0,1)$.
	\qed
\end{prop}
{The proof uses Proposition \ref{prop:positive_drift_of_M_{hh'}(n)} and the fact that policy $\pi^{\star}(L,\delta)$ stops a.s. in finite time. Further, the average in the numerator of the modified GLR statistic, in place of the maximum in the classical GLR statistic, plays a role. For details, see Section \ref{appndx:pi_{LRMB}(L,delta)_belongs_to_Pi(epsilon)}.}

\subsubsection{Upper Bound on the Expected Stopping Time of Policy $\pi^\star(L,\delta)$}
We conclude this section by presenting an upper bound on the expected stopping time of the policy $\pi^\star(L,\delta)$. We show that this upper bound may be made arbitrarily close to the lower bound in \eqref{eq:lower_bound} by tuning $\delta$ appropriately.

As a first step, we show that under the non-stopping version of policy $\pi^\star(L,\delta)$, when $C=(h,P_1,P_2)$ is the underlying configuration of the arms, the modified GLR process has an asymptotic drift that is close to $D^*(h,P_1,P_2)$ that appears in the lower bound \eqref{eq:lower_bound}.

\begin{prop}\label{prop:lim_M_h(n)/n_correct_drift}
	Let $C=(h,P_1,P_2)$ denote the underlying configuration. Fix  $L\geq 1$ and $\delta\in(0,1)$. Then, under the non-stopping version of policy $\pi^{\star}(L,\delta)$, for any $h'\neq h$, we have
	\begin{equation}
		\lim\limits_{n\to\infty}\frac{M_{hh'}(n)}{n}=D_\delta^*(h,P_1,P_2) \quad a.s.,\label{eq:lim_M_{h}(n)/n_almost_correct_drift}
	\end{equation}
	where the quantity $D_\delta^*(h,P_1,P_2)$ is given by
\begingroup\allowdisplaybreaks\begin{align}
  D_\delta^*(h,P_1,P_2)=\lambda_\delta^*\,D(P_1||P_\delta|\mu_1)
		+(1-\lambda_\delta^*)\frac{(K-2)}{(K-1)}D(P_2||P_\delta|\mu_2),\label{eq:L_delta^*(h,P_1,P_2)}
    \end{align}\endgroup
    with $\lambda_\delta^*=\frac{\delta}{K}+(1-\delta)\lambda^*\in [0,1]$,
    and for each $i,j\in\mathcal{S}$, $P_\delta(j|i)$ is as in \eqref{eq:P(j|i)} with $\lambda_1$ replaced by $\lambda_\delta^*$.
    \qed
\end{prop}
{We note that the policy $\pi^*(L,\delta)$ works with only estimated $\hat{P}^n_{h^*(n),1}$ and $\hat{P}^n_{h^*(n),2}$. To show \eqref{eq:lim_M_{h}(n)/n_almost_correct_drift} , we must therefore ensure that the estimates approach the true values and a property akin to continuity holds, that is, taking actions based on $\hat{P}^n_{h^*(n),1}$ and $\hat{P}^n_{h^*(n),2}$, which are only approximately close to $P_1$ and $P_2$, adds only $o(1)$ to the drift $D_\delta^*(h,P_1,P_2)$. This is the notion of certainty equivalence in control theory. The details of the proof may be found in Section \ref{appndx:proof_of_prop_lim_M_h(n)/n_correct_drift}.}

We now state the main result of this section.
\begin{prop}\label{prop:upper_bound}
	Let $C=(h,P_1,P_2)$ denote the underlying configuration of the arms. Fix parameters $L\geq 1$ and $\delta\in(0,1)$. Then, under policy $\pi=\pi^\star(L,\delta)$, we have
	\begin{equation}
		\limsup\limits_{L\to\infty}\frac{E^\pi[\tau(\pi)|C]}{\log L}\leq \frac{1}{D_\delta^*(h,P_1,P_2)}.\label{eq:upper_bound}
	\end{equation}
	\qed
\end{prop}
{The proof uses Proposition \ref{prop:lim_M_h(n)/n_correct_drift} and involves showing that (a) the stopping time $\tau(\pi)$ satisfies an asymptotic almost sure upper bound that matches with the right-hand side of \eqref{eq:upper_bound}, and (b) the family $\{\tau(\pi^\star(L,\delta))/\log L:L\geq 1\}$ is uniformly integrable. The almost sure convergence together with uniform integrability then yields the relation \eqref{eq:upper_bound}. The details may be found in Section \ref{appndx:proof_of_upper_bound}.}

It is clear that $D^*_\delta(h,P_1,P_2)$ is a continuous function of $\delta$, with the property that
\begin{equation}
	\lim\limits_{\delta\downarrow 0}D_\delta^*(h,P_1,P_2)=D^*(h,P_1,P_2),\label{eq:L_delta^*(h,P_1,P_2)_converges_to_D^*(h,P_1,P_2)_as_delta_goes_to_0}
\end{equation}
where $D^*(h,P_1,P_2)$ on the right-hand side of \eqref{eq:L_delta^*(h,P_1,P_2)_converges_to_D^*(h,P_1,P_2)_as_delta_goes_to_0} is the same the constant that appears in the lower bound of \eqref{eq:lower_bound}. Thus, we note that $\delta$ may be tuned to make $D_\delta^*(h,P_1,P_2)$ as close as desired to $D^*(h,P_1,P_2)$, hence establishing the near-optimality of the policy $\pi^\star(L,\delta)$.

\section{The Main Result}\label{sec:main_result}
We now present the main result of this paper, combining the lower and upper bounds stated in Section \ref{sec:lower_bound} and Section \ref{sec:achievability} respectively.
\begin{thrm}
	Consider $K\geq 3$ independent Markov processes on a common finite state space that are irreducible, aperiodic and time homogeneous. Suppose that $C=(h,P_1,P_2)$ is the underlying configuration of the arms, where $h$ denotes the index of the odd arm, and $P_2\neq P_1$. Let $(\epsilon_n)_{n\geq 1}$ denote a sequence of error probability values with the property that $\epsilon_n\to 0$ as $n\to\infty$. Then, for each $n$ and $\delta\in(0,1)$, the policy $\pi^\star(L_n,\delta)$ with $L_n=1/\epsilon_n$ belongs to the family $\Pi(\epsilon_n)$. Furthermore, we have
	\begingroup\allowdisplaybreaks\begin{align}
		\liminf\limits_{n\to\infty}\inf\limits_{\pi\in\Pi(\epsilon_n)}\frac{E[\tau(\pi)|C]}{\log L_n}
		=\lim\limits_{\delta\downarrow 0}\lim\limits_{n\to\infty}\frac{E[\tau(\pi^\star(L_n,\delta))|C]}{\log L_n}=\frac{1}{D^*(h,P_1,P_2)}.
	\end{align}\endgroup
	\qed
\end{thrm}
\begin{IEEEproof}
From Proposition \ref{prop:lower_bound}, it follows that the expected stopping time of any policy $\pi\in\Pi(\epsilon_n)$ grows as $\frac{\log L_n}{D^*(h,P_1,P_2)}$ for large values of $n$. Also, from Proposition \ref{prop:pi_{LRMB}(L,delta)_belongs_to_Pi(epsilon)}, policy $\pi^\star(L_n,\delta)$ belongs to the family $\Pi(\epsilon_n)$ and, from Proposition \ref{prop:upper_bound}, achieves an asymptotic growth of at most $(\log L_n)/D_\delta^*(h,P_1,P_2)$. Since $\lim\limits_{\delta\downarrow 0}D_\delta^*(h,P_1,P_2)=D^*(h,P_1,P_2)$, we may approach the lower bound by choosing an arbitrarily small value of $\delta$. This establishes the theorem.
\end{IEEEproof}

{While those familiar with such stopping problems may easily guess the form of $D^*(h,P_1,P_2)$, the proof is not a straighforward extension of the iid case. To re-emphasise the challenges posed by the setting of Markov rewards, Wald's identity is not available for the converse and a generalisation is needed, while a forced exploration approach provides achievability.}

\section{Simulation Results}\label{sec:simulation_results}
\begin{figure}
	\begin{center}
		\includegraphics[width=11cm,height=7cm]{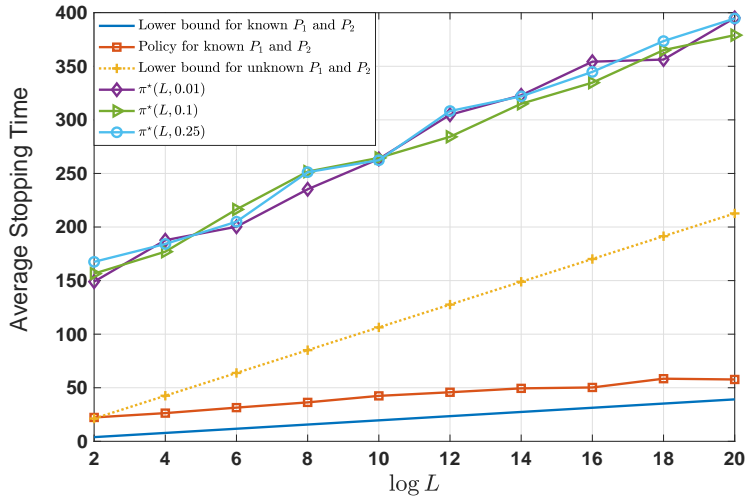}
	\end{center}
	\caption{Plots of average stopping time of policy $\pi^{\star}(L,\delta)$, as function of $\log L$, for $\delta=0.01,0.1,0.25$.}
	\label{fig:plots}
\end{figure}
Fix $K=8$ and $C=(h,P_1,P_2)$, with $h=1$ and $$ P_1=\begin{bmatrix}
	0.5&0.5\\0.5&0.5
\end{bmatrix},\quad  P_2=\begin{bmatrix}
	0.1&0.9\\0.9&0.1
\end{bmatrix}.$$
Fig. \ref{fig:plots} depicts the average stopping time of policy $\pi^{\star}(L,\delta)$ as a function of $\log L$, averaged over $100$ rounds of iterations, for $\delta=0.01,0.1,0.25$. For the aforementioned values of $P_1$ and $P_2$, numerical evaluation yields $D^*(h,P_1,P_2) \simeq 0.094$, thus resulting in a lower bound of $1/D^*(h,P_1,P_2) \simeq 10.635$. Since \eqref{eq:lower_bound} is a statement about the slope of the growth rate of average stopping time of policy $\pi^{\star}(L,\delta)$ as a function of $\log L$, the top 3 plots in the figure respect the lower bound in \eqref{eq:lower_bound}, with the slopes in these plots only marginally higher than that given by the lower bound. Theory predicts that as $\delta \downarrow 0$ and $L\to\infty$, the slopes will approach the lower bound. Also included in the figure are the plots of (a) the lower bound for the case when $P_1$ and $P_2$ are known, and (b) a policy similar to that of $\pi^{\star}(L,\delta)$ that uses the knowledge of $P_1$ and $P_2$ to identify the index of the odd arm. Such a policy clearly takes lesser time than $\pi^\star(L,\delta)$ to identify the index of the odd arm. The figure shows that the performance of this policy also matches in slope to that given by its lower bound for large values of $L$.

\section{Proofs of the Main Results}\label{sec:proofs_of_main_results}
\subsection{Proof of Proposition \ref{prop:lower_bound}}\label{appndx:proof_of_lower_bound}
We first present below 3 lemmas that will be used in the proof of the proposition. The first of these, given below, is an analogue of the change of measure argument of Kaufmann et al. \cite[Lemma 18]{kaufmann2016complexity} for the case of Markov observations from each arm.

Recall the definition of $\mathcal{F}_\tau$ in \eqref{eq:stopping_time_sigma_alg}:
	\begin{equation*}
	\mathcal{F}_\tau=\{E\in\mathcal{F}:E\cap \{\tau=n\}\in\mathcal{F}_n\text{ for all }n\geq 0 \},	\end{equation*}
	where for each $n$, $\mathcal{F}_n$ is as defined in \eqref{eq:filtration}. Further, for any $h'\neq h$, define $Z_{hh'}(\tau)\coloneqq Z_{h}(\tau)-Z_{h'}(\tau)$, where $Z_{h}(\tau)=\sum_{a=1}^{K}Z_{h}^a(\tau)$.

\begin{lemma}\label{lemma:ChangeOfMeasure_htoh'}
	Fix $\epsilon>0$ and probability transition matrices $P_1$ and $P_2$, and let $\tau$ be the stopping time of a policy $\pi\in\Pi(\epsilon)$. Then, for any event $E\in \mathcal{F}_{\tau}$ and configuration triplets $C=(h,P_1,P_2)$ and $C'=(h',P_1',P_2')$, with $h'\neq h$, we have
	\begin{equation}
	P^\pi(E|C')=E^\pi[1_{E}\,\,\exp(-Z_{hh'}(\tau))|C].\label{eq:statement_of_lemma_ChangeOfMeasure_htoh'}
	\end{equation}
	\qed
\end{lemma}

\begin{IEEEproof}
The proof follows the outline in \cite{kaufmann2016complexity}, with crucial modifications needed for the Markov problem at hand. We use the shorthand notations $E_{h}[\cdot]$ and $E_{h'}[\cdot]$ to denote respectively the quantities $E^\pi[\cdot|C]$ and $E^\pi[\cdot|C']$; similarly, $P_{h}(\cdot)$ and $P_{h'}(\cdot)$ denote the respective probabilities. We begin by showing that for all $n\geq 0$, the following statement is true: for any measurable function $g:\mathcal{A}^{n+1}\times\mathcal{S}^{n+1}\to \mathbb{R}$, we have
\begin{equation}
E_{h'}[g(A^n,\bar{X}^{n})]=E_{h}[g(A^n,\bar{X}^{n})\,\exp(-Z_{hh'}(n))].\label{eq:for_all_g_mble_lemma}
\end{equation}
Assuming that the above statement is true, for any $E\in \mathcal{F}_{\tau}$, we have
\begingroup\allowdisplaybreaks\begin{align}
P_{h'}(E)&=E_{h'}[1_{E}]\nonumber\\
&\stackrel{(a)}{=}\sum\limits_{n=0}^{\infty} E_{h'}[1_{E}1_{\{\tau=n\}}]\nonumber\\
&\stackrel{(b)}{=}\sum\limits_{n=0}^{\infty} E_{h}[1_{E}1_{\{\tau=n\}}\,\exp(-Z_{hh'}(n))]\nonumber\\
&=E_{h}[1_{E}\,\,\exp(-Z_{hh'}(\tau))],\label{eq:proof_lem_1}
\end{align}\endgroup
hence proving the lemma. In the above set of equations, $(a)$ is due to monotone convergence theorem, and $(b)$ follows from the application of \eqref{eq:for_all_g_mble_lemma} to the function $g(A^n,\bar{X}^{n})=1_{E}\cdot1_{\{\tau=n\}}$ by noting that $E\in \mathcal{F}_{\tau}$, and therefore $E\cap\{\tau=n\}\in \mathcal{F}_{n}$ for all $n$.

We now proceed to prove \eqref{eq:for_all_g_mble_lemma} by induction on $n$.
From \eqref{eq:log_likelihood_under_hyp_h_and_h'} and  \eqref{eq:Z_{hh'}^a(n)}, it follows that $Z_{hh'}(0)=0$. Then, for any measurable function $g:\mathcal{A}^{n+1}\times \mathcal{S}^{n+1}\to\mathbb{R}$, the proof of \eqref{eq:for_all_g_mble_lemma} for $n=0$ follows from the following set of equations.
\begingroup\allowdisplaybreaks\begin{align}
	E_{h'}[g(A_0,\bar{X}_0)]
	&=\sum\limits_{a=1}^{K}\sum\limits_{i\in\mathcal{S}}P_{h'}(A_0=a)\cdot P_{h'}(\bar{X}_0=i|A_0=a)\cdot g(a,i)\nonumber\\
	&=\sum\limits_{a=1}^{K}\sum\limits_{i\in\mathcal{S}}P_{h'}(A_0=a)\cdot P_{h'}(X_0^a=i)\cdot g(a,i)\nonumber\\
	&=\sum\limits_{a=1}^{K}\sum\limits_{i\in\mathcal{S}}P_{h'}(A_0=a)\cdot \nu(i)\cdot g(a,i)\nonumber\\
	&\stackrel{(a)}{=}\sum\limits_{a=1}^{K}\sum\limits_{i\in\mathcal{S}}P_{h}(A_0=a)\cdot P_h(X_0^a=i)\cdot g(a,i)\nonumber\\
	&=E_h[g(A_0,\bar{X}_0)]\nonumber\\
	&=E_h[g(A_0,\bar{X}_0)\exp(-Z_{hh'}(0))],
\end{align}\endgroup
where in writing $(a)$, we use
\begin{itemize}
\item the fact that $P_h(A_0=a)=P_{h'}(A_0=a)$ since the manner in which $A_0$ is selected is not a function of either $h$ or $h'$. For instance, we may assume that each of the arms is picked once in the first $K$ time instants, and note that this does not affect the asymptotic performance of the policy. In such a case, $P_h(A_0=1)=1=P_{h'}(A_0=1)$.
\item the fact that $X_0^a\sim \nu$ under hypotheses $H_h$ and $H_{h'}$.
\end{itemize}

We now assume that \eqref{eq:for_all_g_mble_lemma} holds for some positive integer $n$, and show that it also holds for $n+1$. We have
\begingroup\allowdisplaybreaks\begin{align}
&E_{h'}[g(A^{n+1},\bar{X}^{n+1})]=E_{h'}\left[E_{h'}\left[g(A^{n+1},\bar{X}^{n+1})|A^n,\bar{X}^n\right]\right].\label{eq:induc_hyp_appl}
\end{align}\endgroup
Since the inner conditional expectation term on the right-hand side of \eqref{eq:induc_hyp_appl} is a measurable function of $(A^n,\bar{X}^n)$, using the induction hypothesis, we get
\begingroup\allowdisplaybreaks\begin{align}
& E_{h'}[g(A^{n+1},\bar{X}^{n+1})]\nonumber\\
& =E_{h}\left[E_{h'}\left[g(A^{n+1},\bar{X}^{n+1})|A^n,\bar{X}^n\right]\,\exp(-Z_{hh'}(n))\right]\nonumber\\
&=\sum\limits_{a^n\in\mathcal{A}^n}\sum\limits_{\bar{x}^n\in \mathcal{S}^{n+1}}P_h(A^n=a^n,\bar{X}^n=\bar{x}^n)\cdot \exp(-z_{hh'}(n))\cdot E_{h'}[g(A^{n+1},\bar{X}^{n+1})|A^n=a^n,\bar{X}^n=\bar{x}^n]\label{eq:lem_1_induc_partial_1},
\end{align}\endgroup
where $z_{hh'}(n)$ denotes the value of $Z_{hh'}(n)$ when $A^n=a^n$ and $\bar{X}^n=\bar{x}^n$. Then, we have
\begingroup\allowdisplaybreaks\begin{align}
&E_{h'}[g(A^{n+1},\bar{X}^{n+1})|A^n=a^n,\bar{X}^n=\bar{x}^n]\nonumber\\
&=\sum\limits_{a'=1}^{K}\sum\limits_{j\in\mathcal{S}}g(a^n,a',\bar{x}^n,j)\cdot P_{h'}(A_{n+1}=a'|A^n=a^n,\bar{X}^n=\bar{x}^n)
\cdot P_{h'}^{a'}(X_{N_{a'}(n)}^{a'}=j|X_{N_{a'}(n)-1}^{a'})\nonumber\\
&=\sum\limits_{a'=1}^{K}\sum\limits_{j\in\mathcal{S}}g(a^n,a',\bar{x}^n,j)\cdot P_{h}(A_{n+1}=a'|A^n=a^n,\bar{X}^n=\bar{x}^n)
\cdot P_{h'}^{a'}(X_{N_{a'}(n)}^{a'}=j|X_{N_{a'}(n)-1}^{a'}),\label{eq:lem_1_induc_partial_2}
\end{align}\endgroup
where in writing the last line above, we use the fact that the probability of selecting an arm at any time, based on the history of past arm selections and observations, is independent of the underlying configuration of the arms, and is thus the same under hypotheses $H_h$ and $H_{h'}$. We now write \eqref{eq:lem_1_induc_partial_2} as
\begingroup\allowdisplaybreaks\begin{align}
&E_{h'}[g(A^{n+1},\bar{X}^{n+1})|A^n=a^n,\bar{X}^n=\bar{x}^n]\nonumber\\
&=\sum\limits_{a'=1}^{K}\sum\limits_{j\in\mathcal{S}}\bigg\lbrace g(a^n,a',\bar{x}^n,j)\cdot P_{h}(A_{n+1}=a'|A^n=a^n,\bar{X}^n=\bar{x}^n)\nonumber\\
&\hspace{6cm}
\cdot\frac{P_{h'}^{a'}(X_{N_{a'}(n)-1}^{a'}=j|X_{N_{a'}(n)-1}^{a'})}{P_{h}^{a'}(X_{N_{a'}(n)}^{a'}=j|X_{N_{a'}(n)-1}^{a'})}\cdot P_{h}^{a'}(X_{N_{a'}(n)}^{a'}=j|X_{N_{a'}(n)-1}^{a'})\bigg\rbrace.\label{eq:lem_1_induc_partial_3}
\end{align}\endgroup
Plugging back \eqref{eq:lem_1_induc_partial_3} in \eqref{eq:lem_1_induc_partial_1}, and using
\begin{equation}
	z_{hh'}(n+1)=z_{hh'}(n)+\log\frac{P_{h}^{a'}(X_{N_{a'}(n)}^{a'}=j|X_{N_{a'}(n)-1}^{a'})}{P_{h'}^{a'}(X_{N_{a'}(n)}^a=j|X_{N_{a'}(n)-1}^{a'})},
\end{equation}
we get
\begingroup\allowdisplaybreaks\begin{align}
&E_{h'}[g(A^{n+1},\bar{X}^{n+1})]\nonumber\\
&=\sum\limits_{a^n\in\mathcal{A}^n}\sum\limits_{\bar{x}^n\in \mathcal{S}^{n+1}}\sum\limits_{a'=1}^{K}\sum\limits_{j\in\mathcal{S}}\bigg\lbrace g(a^n,a',\bar{x}^n,j)\cdot \exp(-z_{hh'}(n+1))\nonumber\\
&\hspace{6cm}\cdot P_h(A^n=a^n,\bar{X}^n=\bar{x}^n)
\cdot P_{h}(A_{n+1}=a',\bar{X}_{n+1}=j|A^n=a^n,\bar{X}^n=\bar{x}^n)\bigg\rbrace\nonumber\\
&=E_h[g(A^{n+1},\bar{X}^{n+1})\exp(-Z_{hh'}(n+1))],
\end{align}\endgroup
hence proving \eqref{eq:statement_of_lemma_ChangeOfMeasure_htoh'} .
%
\end{IEEEproof}

The second lemma below relates the expected number of $i$ to $j$ transitions $E^\pi[N_a(\tau,i,j)|C]$ observed on the Markov process of arm $a$ to $E^\pi[N_a(\tau,i)|C]$, the expected number of exits out of state $i$ observed on the Markov process of arm $a$.

\begin{lemma}\label{lemma:RelBtwNijAndNi}
	Fix $\epsilon>0$, a policy $\pi\in\Pi(\epsilon)$, and a configuration $C=(h,P_1,P_2)$. For each $i,j\in \mathcal{S}$ and $a\in \mathcal{A}$, we have
	\begin{equation}
	E^\pi[N_a(\tau,i,j)|C]=E^\pi[N_a(\tau,i)|C]\cdot P_{h}^a(j|i),\label{eq:RelBtwNijAndNi}
	\end{equation}
	where $P_h^a(j|i)$ is as given in \eqref{eq:P_h^a(j|i)}.
	\qed
\end{lemma}

\begin{IEEEproof}
We use the shorthand notation $E_h[\cdot]$ to denote $E^\pi[\cdot|C]$.
We demonstrate that for each $i,j\in \mathcal{S}$ and $a\in \mathcal{A}$,
\begingroup\allowdisplaybreaks\begin{align}
{E}_{h}[{E}_{h}[N_a(\tau,i,j)|X_0^a]|N_a(\tau)]
=E_{h}[E_{h}[N_a(\tau,i)|X_0^a]|N_a(\tau)]\cdot P_h^a(j|i).\label{eq:RelBtwNijAndNiWithIteratedExpec}
\end{align}\endgroup
Towards this, we note that
\begingroup\allowdisplaybreaks\begin{align}
E_{h}[E_{h}[N_a(\tau,i,j)|X_0^a]|N_a(\tau)]
=E_{h}\left[\sum\limits_{m=1}^{N_a(\tau)-1}E_{h}[1_{\{X_{m-1}^a=i,\,X_m^a=j\}}|X_0^a]\bigg\vert N_a(\tau)\right].\label{eq:lower_bound_partial_5_rested_arms}
\end{align}\endgroup
We now simplify the inner conditional expectation term in \eqref{eq:lower_bound_partial_5_rested_arms} by considering the cases $m=1$ and $m\geq 2$ separately.
\begin{enumerate}
	\item Case $m=1$:
	In this case, we get
	\begingroup\allowdisplaybreaks\begin{align}
	E_{h}[1_{\{X_{0}^a=i,\,X_1^a=j\}}|X_0^a]
	&=1_{\{X_{0}^a=i\}}\cdot E_{h}[1_{\{X_1^a=j\}}|X_0^a]\nonumber\\
	&=1_{\{X_{0}^a=i\}}\cdot P_h^a(X_1^a=j|X_0^a=i)\nonumber\\
	&=1_{\{X_{0}^a=i\}}\cdot P_h^a(j|i).\label{eq:Casem=1}
	\end{align}\endgroup
	\item Case $m\geq 2$: Here, we get
	\begingroup\allowdisplaybreaks\begin{align}
	E_{h}[1_{\{X_{m-1}^a=i,\,X_m^a=j\}}|X_0^a=k]
	&=P_h^a(X_{m-1}^a=i,\,X_m^a=j|X_0^a=k)\nonumber\\
	&\stackrel{(a)}{=}P_h^a(X_{m-1}^a=i|X_0^a=k)\cdot P_h^a(X_1^a=j|X_0^a=i)\nonumber\\
	&=E_{h}[1_{\{X_{m-1}^{a}=i\}}|X_0^a=k]\cdot P_h^a(j|i),
	\end{align}\endgroup
	from which it follows that $E_{h}[1_{\{X_{m-1}^a=i,\,X_m^a=j\}}|X_0^a]=E_{h}[1_{\{X_{m-1}^{a}=i\}}|X_0^a]\cdot P_h^a(j|i)$. In the above set of equations, $(a)$ follows from the fact that the Markov process of arm $a$ is time homogeneous.
\end{enumerate}
From the aforementioned cases, it follows that the relation
\begingroup\allowdisplaybreaks\begin{align}
E_{h}[1_{\{X_{m-1}^a=i,\,X_m^a=j\}}|X_0^a]=E_{h}[1_{\{X_{m-1}^{a}=i\}}|X_0^a]\cdot P_h^a(j|i)\label{eq:ConditioningOnX_0^a}
\end{align}\endgroup
holds for all $m\geq 1$. Substituting \eqref{eq:ConditioningOnX_0^a} in \eqref{eq:lower_bound_partial_5_rested_arms} and simplifying, we arrive at \eqref{eq:RelBtwNijAndNiWithIteratedExpec}. The lemma then follows by applying expectation $E_{h}[\cdot]$ to both sides of \eqref{eq:RelBtwNijAndNiWithIteratedExpec}.	
\end{IEEEproof}

The third lemma presented below will be used to simplify a minimisation term later in the proof of the proposition.
 \begin{lemma}
 	Denote by $\mathcal{P}(\mathcal{S})$ the set of all probability distributions on the set $\mathcal{S}$, and let $\nu_1$ and $\nu_2$ be any two distinct elements of $\mathcal{P}(\mathcal{S})$. Then, for all $w_1,w_2\in[0,1]$ such that $w_1+w_2=1$, we have
 	\begingroup\allowdisplaybreaks\begin{align}
 		\min\limits_{\psi\in\mathcal{P}(\mathcal{S})}\left[w_1 D(\nu_1||\psi)+w_2 D(\nu_2||\psi)\right]
 		=w_1 D(\nu_1||\nu^*)+w_2 D(\nu_2||\nu^*),\label{eq:convex_comb_of_KL_div}
 	\end{align}\endgroup
 	where $\nu^*\in\mathcal{P}(\mathcal{S})$ is given by $\nu^*=w_1\nu_1+w_2\nu_2$.
 	\qed
 \end{lemma}
 \begin{IEEEproof}
 This is well known with $\nu^*$ viewed as a root of ``information centre'' and the right-hand side of \eqref{eq:convex_comb_of_KL_div} viewed as a mutual information. Here is the proof for completeness.

  Let $\nu^*$ be as defined in the statement of the lemma. For any $\psi\in\mathcal{P}(\mathcal{S})$, we have
{\color{black}
 \begingroup\allowdisplaybreaks\begin{align}
 w_1 D(\nu_1||\psi)+w_2 D(\nu_2||\psi)
 &=w_1D(\nu_1\|\nu^*)+w_2 D(\nu_2\| \nu^*)+D(\nu^*||\psi)\nonumber\\
 &\geq D(\nu_1\|\nu^*)+w_2 D(\nu_2\| \nu^*),\label{eq:conv_comb_of_KL_div_proof}
 \end{align}\endgroup}
 with equality in the last line above if and only if $\psi=\nu^*$. This completes the proof of the lemma.	
 \end{IEEEproof}

\begin{IEEEproof}[Proof of Proposition \ref{prop:lower_bound}]
Fix an arbitrary $\epsilon>0$, and let $\pi\in\Pi(\epsilon)$ be a policy whose stopping is $\tau=\tau(\pi)$. Without loss of generality, we assume that $E^\pi[\tau(\pi)|C] < \infty$, for otherwise the inequality \eqref{eq:lower_bound} holds trivially. We organise the proof of the proposition into various sections. In the first of these sections presented below, we lower bound the expected value of $Z_{hh'}(\tau)$ in terms of the error probability $\epsilon$. This uses the above Lemma \ref{lemma:ChangeOfMeasure_htoh'}, Lemma \ref{lemma:RelBtwNijAndNi} and the result of \cite[Lemma 19]{kaufmann2016complexity}.

\subsubsection{A Lower Bound on The Expected Value of $Z_{hh'}(\tau)$}
Let $\pi\in\Pi(\epsilon)$, with stopping time is $\tau=\tau(\pi)$. For any $h'\neq h$, let $Z_{hh'}(\tau)$ be as defined in the statement of Lemma \ref{lemma:ChangeOfMeasure_htoh'}. Then, Lemma \ref{lemma:ChangeOfMeasure_htoh'} in conjunction with \cite[Lemma 19]{kaufmann2016complexity} yields the following: conditioned on the underlying configuration $C=(h,P_1,P_2)$, for any alternative configuration $C'=(h',P_1',P_2')$, where $h'\neq h$, under the assumption that $E^\pi[\tau|C]<\infty$, we have
\begin{equation}
E^\pi[Z_{hh'}(\tau)|C]\geq \sup\limits_{E\in \mathcal{F}_{\tau}}d(P^\pi(E|C),P^\pi(E|C')),\label{eq:analog_of_lemma_19_Kaufmann}
\end{equation}
 where \[d(p,q)\coloneqq p\log\left(\frac{p}{q}\right)+(1-p)\log\left(\frac{1-p}{1-q}\right)\]
denotes the binary KL divergence, with the convention that $d(0,0)=0=d(1,1)$. We now note the following points:
\begin{enumerate}
    \item For each alternative configuration $C'$, by taking $E=\{I(\pi)=h\}$ and recognising that $\pi\in\Pi(\epsilon)$, we have $P^\pi(E|C)>1-\epsilon$ and $P^\pi(E|C')\leq \epsilon$. Using this, along with the fact that the mapping $x\mapsto d(x,y)$ is monotone increasing for $x<y$ and the mapping $y\mapsto d(x,y)$ is monotone decreasing for any fixed $x$, we obtain
	\begingroup\allowdisplaybreaks\begin{align}
		d(P^\pi(E|C),P^\pi(E|C'))&\geq d(1-\epsilon,P^\pi(E|C'))\nonumber\\
		&\geq d(1-\epsilon,\epsilon).
	\end{align}\endgroup
	
	\item We may minimise both sides of \eqref{eq:analog_of_lemma_19_Kaufmann} over all alternative configurations $C'$ to obtain
	\begingroup\allowdisplaybreaks\begin{align}
		\min\limits_{C'=(h',P_1',P_2')}E^\pi[Z_{hh'}(\tau)|C]
		\geq \min\limits_{C'=(h',P_1',P_2')}\,\sup\limits_{E\in \mathcal{F}_{\tau}}d(P^\pi(E|C),P^\pi(E|C')).\label{eq:analog_of_lemma_19_Kaufmann_min_over_all_C'}
	\end{align}\endgroup
\end{enumerate}

Combining the points noted above, and using $d(1-\epsilon,\epsilon)=d(\epsilon,1-\epsilon)$, we obtain
\begin{equation}
\min\limits_{C'=(h',P_1',P_2')}{E}^\pi[Z_{hh'}(\tau)|C]\geq d(\epsilon,1-\epsilon).\label{eq:lower_bound_partial_1}
\end{equation}

\subsubsection{A Relation Between $E^\pi[Z_{hh'}(\tau)|C]$ and $E^\pi[\tau|C]$}
 As our next step, we obtain an upper bound for $E^\pi[Z_{hh'}(\tau)|C]$ in terms of $E^\pi[\tau|C]$. Towards this, we have
\begingroup\allowdisplaybreaks\begin{align}
{E}^\pi[Z_{hh'}(\tau)|C]
=\sum\limits_{a=1}^{K}{E}^\pi\bigg[\sum\limits_{m=1}^{N_{a}(\tau)-1}\log\left(\frac{P_{h}^{a}(X_{m}^{a}|X_{m-1}^{a})}{P_{h'}^{a}(X_{m}^{a}|X_{m-1}^{a})}\right)\bigg\vert C\bigg],\label{eq:lower_bound_partial_2_rested}
\end{align}\endgroup
where we take inner summation term to be zero whenever $N_a(\tau)<2$.
Focus on the expectation term in \eqref{eq:lower_bound_partial_2_rested}.
This term may be written as
\begingroup\allowdisplaybreaks\begin{align}
E^\pi\bigg[\sum\limits_{m=1}^{N_{a}(\tau)-1}\log\left(\frac{P_{h}^{a}(X_{m}^{a}|X_{m-1}^{a})}{P_{h'}^{a}(X_{m}^{a}|X_{m-1}^{a})}\right)\bigg\vert C\bigg]
&\stackrel{(a)}{=}E^\pi\bigg[\sum\limits_{m=1}^{N_{a}(\tau)-1}\sum\limits_{i,j\in S}1_{\{X_{m-1}^a=i,\,X_m^a=j\}}\log\left(\frac{P_{h}^{a}(j|i)}{P_{h'}^{a}(j|i)}\right)\bigg\vert C\bigg]\nonumber\\
&=\sum\limits_{i,j\in S}E^\pi[N_a(\tau,i,j)|C]\,f^a_{hh'}(j|i),\label{eq:lower_bound_partial_3_rested_bandit}
\end{align}\endgroup
where $(a)$ above follows from the fact that the Markov process of arm $a$ is time homogeneous, and $f_{hh'}^a(j|i)\coloneqq\log\left(\frac{P_{h}^{a}(j|i)}{P_{h'}^{a}(j|i)}\right)$. Using the result of Lemma \ref{lemma:RelBtwNijAndNi} in \eqref{eq:lower_bound_partial_3_rested_bandit}, we get
\begingroup\allowdisplaybreaks\begin{align}
E^\pi[Z_{hh'}(\tau)|C]
&=\sum\limits_{a=1}^{K}\,\sum\limits_{i,j\in S}{E}^\pi[N_a(\tau,i)|C]\cdot P_{h}^a(j|i)\cdot f^a_{hh'}(j|i)\nonumber\\
&=\sum\limits_{a=1}^{K}\,\sum\limits_{i\in S}{E}[N_a(\tau,i)|C]\,D(P_h^a(\cdot|i)||P_{h'}^a(\cdot|i)),\label{eq:lower_bound_partial_6_rested_arms}
\end{align}\endgroup
where $D(P_h^a(\cdot|i)||P_{h'}^a(\cdot|i))=\sum\limits_{j\in S}P_{h}^a(j|i)f^a_{hh'}(j|i)$ denotes the KL divergence between the probability distributions $P_h^a(\cdot|i)$ and $P_{h'}^a(\cdot|i)$. We now express \eqref{eq:lower_bound_partial_6_rested_arms} by introducing some additional terms as below:
\begingroup\allowdisplaybreaks\begin{align}
&E^\pi[Z_{hh'}(\tau)|C]\nonumber\\
&=(E^\pi[\tau+1|C]-K)
\bigg(\sum\limits_{a=1}^{K}\bigg[\frac{E^\pi[N_a(\tau)|C]-1}{E^\pi[\tau+1|C]-K}\bigg]\sum\limits_{i\in \mathcal{S}}\bigg[\frac{E^\pi[N_a(\tau,i)|C]}{E^\pi[N_a(\tau)|C]-1}\bigg]
D(P_h^a(\cdot|i)||P_{h'}^a(\cdot|i))\bigg)\nonumber\\
&=(E^\pi[\tau+1|C]-K)\bigg(\sum\limits_{a=1}^{K}\bigg[\frac{E^\pi[N_a(\tau)|C]-1}{E^\pi[\tau+1|C]-K}\bigg]
\sum\limits_{i\in \mathcal{S}}p_h^a(i)\cdot D(P_h^a(\cdot|i)||P_{h'}^a(\cdot|i))\bigg),\label{eq:lower_bound_partial_7_rested_arms}
\end{align}\endgroup
where $p_h^a(i)\coloneqq\frac{E^\pi[N_a(\tau,i)|C]}{E^\pi[N_a(\tau)|C]-1}$ represents the average (computed with respect to $E^\pi[\cdot|C])$ fraction of times a transition {out} of state $i$ is observed on the Markov process of arm $a$.

\subsubsection{Asymptotics of Vanishing Error Probability}
Since $\sum\limits_{i\in \mathcal{S}}p_h^a(i)=1$, the inner summation term over $i$ in \eqref{eq:lower_bound_partial_7_rested_arms} represents the average of the numbers $(D(P_h^a(\cdot|i)||P_{h'}^a(\cdot|i)))_{i\in \mathcal{S}}$ with respect to $(p_h^a(i))_{i\in \mathcal{S}}$. Suppose that at some time, arm $a$ is selected, and it \textcolor{black}{makes a transition from state $i$ to state $j$}, for some $i,j\in \mathcal{S}$. Then, the next time arm $a$ is selected, it \textcolor{black}{makes a transition from state $j$ to state $k$} for some $k\in \mathcal{S}$. For $a\in \mathcal{A}$ and $i\in \mathcal{S}$, let
\begin{equation}
N^a(\tau,i)\coloneqq\sum\limits_{m=2}^{N_a(\tau)}1_{\{X_{m-1}^a=i\}}\label{eq:NoOfEntriesIntoStatei}
\end{equation}
denote the number of times arm $a$ \textcolor{black}{is observed to occupy state $i$ after a transition}. In conjunction with \eqref{eq:N_a(n,i)}, it is easy to see that for each $i\in \mathcal{S}$, we have
\begin{equation}
N_a(\tau,i)=N^a(\tau,i)-1_{\{X_{N_a(\tau)-1}^a=i\}}+1_{\{X_0^a=i\}},\label{eq:RelBtwEntryAndExitNumbersForStatei}
\end{equation}
which implies that $N^a(\tau,i)-1\leq N_a(\tau,i)\leq N^a(\tau,i)+1$ a.s. Thus, we notice that for the Markov process of each arm, for each $i\in\mathcal{S}$, \textcolor{black}{the number of times the arm is observed to occupy state $i$ prior to a transition is at most one more than the number of times it is observed to occupy state $i$ after a transition}.  We then have
\begin{equation}
\frac{E^\pi[N^a(\tau,i)|C]-1}{E^\pi[N_a(\tau)|C]-1}\leq p_h^a(i)\leq \frac{E^\pi[N^a(\tau,i)|C]+1}{E^\pi[N_a(\tau)|C]-1}.\label{eq:p_h^a_in_limit}
\end{equation}

Using \eqref{eq:p_h^a_in_limit} in \eqref{eq:lower_bound_partial_7_rested_arms}, we arrive at the form
\begingroup\allowdisplaybreaks\begin{align}
	u-\Delta\leq E^\pi[Z_{hh'}(\tau)]
	\leq u+\Delta,\label{eq:lower_bound_partial_8}
\end{align}\endgroup
where the terms $u$ and $\Delta$ are as below:
\begingroup\allowdisplaybreaks\begin{align}
u&=(E^\pi[\tau+1|C]-K)
\bigg(\sum\limits_{a=1}^{K}\bigg[\frac{E^\pi[N_a(\tau)|C]-1}{E^\pi[\tau+1|C]-K}\bigg]
\sum\limits_{i\in \mathcal{S}}\bigg[\frac{E^\pi[N^a(\tau,i)|C]}{E^\pi[N_a(\tau)|C]-1}\bigg] D(P_h^a(\cdot|i)||P_{h'}^a(\cdot|i))\bigg),\nonumber\\
\Delta &=\sum\limits_{a=1}^{K}\sum\limits_{i\in\mathcal{S}}D(P_h^a(\cdot|i)||P_{h'}^a(\cdot|i))
=\sum\limits_{i\in\mathcal{S}}D(P_1(\cdot|i)||P_2'(\cdot|i))+\sum\limits_{i\in\mathcal{S}}D(P_2(\cdot|i)||P_1'(\cdot|i))
+\sum\limits_{a\neq h}\sum\limits_{i\in\mathcal{S}}D(P_2(\cdot|i)||P_2'(\cdot|i)).
\end{align}\endgroup

We shall soon show that the regime of vanishing error probabilities, i.e., $\epsilon\downarrow 0$, necessarily means that for each $a\in\mathcal{A}$, $E^\pi[N_a(\tau)|C]\to \infty$, which in turn implies that $E^\pi[\tau|C]\to\infty$. In this asymptotic regime, for each $a\in\mathcal{A}$, the limiting probabilities of \textcolor{black}{arm $a$ occupying a state $i\in \mathcal{S}$ prior to and after a transition are equal, and invariant to the one step transitions on arm $a$}. Since the Markov process of arm $a$ is irreducible and positive recurrent, its probability transition matrix admits a unique stationary distribution. Therefore, by the Ergodic theorem, the aforementioned probabilities must converge to those given by the stationary distribution associated with arm $a$. We shall denote this stationary distribution by $\mu_h^a(\cdot)$ under configuration $C=(h,P_1,P_2)$, given by
\begin{equation}
	\mu_h^a(i)=\begin{cases}
		\mu_1(i),&a=h,\\
		\mu_2(i),&a\neq h.
	\end{cases}\label{eq:mu_h^a}
\end{equation}
Then, as $\epsilon\downarrow 0$, we have that both the lower and upper bounds in \eqref{eq:p_h^a_in_limit} converge to $\mu_h^a(i)$ . We shall soon exploit this fact below to arrive at the lower bound. Going further, we denote by $(q_h^a(i))_{i\in\mathcal{S}}$ the probability distribution given by
\begin{equation}
q_h^a(i)=\frac{E^\pi[N^a(\tau,i)|C]}{E^\pi[N_a(\tau)|C]-1},\quad i\in\mathcal{S}.\label{eq:q_h^a}	
\end{equation}

Using the upper bound in \eqref{eq:lower_bound_partial_8} in combination with \eqref{eq:lower_bound_partial_1}, we have the following chain of inequalities:
\begingroup\allowdisplaybreaks\begin{align}
d(\epsilon,1-\epsilon)
&\leq\min\limits_{C'=(h',P_1',P_2')}\,{E}^\pi[Z_{hh'}(\tau)|C]\nonumber\\
&\leq \min\limits_{C'=(h',P_1',P_2')}(u+\Delta)\nonumber\\
&\leq \min\limits_{C'=(h',P_1',P_2')} u+\min\limits_{C'=(h',P_1',P_2')} \Delta.\label{eq:lower_bound_partial_9_rested_arms}
\end{align}\endgroup

The first term in \eqref{eq:lower_bound_partial_9_rested_arms} may be upper bounded as follows:
{\color{black}
\begingroup\allowdisplaybreaks\begin{align}
\min\limits_{C'=(h',P_1',P_2')}u &	
= (E^\pi[\tau+1|C]-K)\,\bigg\lbrace
\min\limits_{C'=(h',P_1',P_2')}\bigg(\sum\limits_{a=1}^{K}\bigg[\frac{E^\pi[N_a(\tau)|C]-1}{E^\pi[\tau+1|C]-K}\bigg]\sum\limits_{i\in\mathcal{S}}q_h^a(i)\, D(P_h^a(\cdot|i)||P_{h'}^a(\cdot|i))\bigg)\bigg\rbrace\nonumber\\
&\stackrel{(a)}{=} (E^\pi[\tau+1|C]-K)\,\bigg\lbrace
\min\limits_{C'=(h',P_1',P_2')}\bigg(\sum\limits_{a=1}^{K}\bigg[\frac{E^\pi[N_a(\tau)|C]-1}{E^\pi[\tau+1|C]-K}\bigg]
 D(P_h^a(\cdot|\cdot)||P_{h'}^a(\cdot|\cdot)|q_h^a)\bigg)\bigg\rbrace\nonumber\\
&\stackrel{(b)}{\leq} (E^\pi[\tau+1|C]-K)\,\bigg\lbrace\max\limits_{\lambda\in\mathcal{P}(\mathcal{A})}\min\limits_{C'=(h',P_1',P_2')}\bigg(\sum\limits_{a=1}^{K}\lambda(a) D(P_h^a(\cdot|\cdot)||P_{h'}^a(\cdot|\cdot)|q_h^a)\bigg)\bigg\rbrace,\label{eq:proof_of_prop_1_temp_1}
\end{align}\endgroup}
where, in $(a)$ above,
\begin{align*}
D(P_h^a(\cdot|\cdot)||P_{h'}^a(\cdot|\cdot)|q_h^a)\coloneqq \sum\limits_{i\in S}q_h^a(i)\cdot D(P_h^a(\cdot|i)||P_{h'}^a(\cdot|i)),
\end{align*}
while $(b)$ follows by noting that maximising over the set $\mathcal{P}(\mathcal{A})$ of all probability distributions on the set of arms $\mathcal{A}$ only increases the right-hand side. The second term in \eqref{eq:lower_bound_partial_9_rested_arms} may be simplified as
\begingroup\allowdisplaybreaks\begin{align}
	\min\limits_{C'=(h',P_1',P_2')}\Delta
	&=\min\limits_{P_1',P_2':P_1'\neq P_2'}\bigg\lbrace \sum\limits_{i\in\mathcal{S}}D(P_1(\cdot|i)||P_2'(\cdot|i))+\sum\limits_{i\in\mathcal{S}}D(P_2(\cdot|i)||P_1'(\cdot|i))
+\sum\limits_{a\neq h}\sum\limits_{i\in\mathcal{S}}D(P_2(\cdot|i)||P_2'(\cdot|i))\bigg\rbrace\nonumber\\
&\stackrel{(a)}{=}\min\limits_{P_2'}\bigg\lbrace \sum\limits_{i\in\mathcal{S}}D(P_1(\cdot|i)||P_2'(\cdot|i))+\sum\limits_{a\neq h}\sum\limits_{i\in\mathcal{S}}D(P_2(\cdot|i)||P_2'(\cdot|i))\bigg\rbrace\nonumber\\
&=\min\bigg\lbrace\sum\limits_{i\in\mathcal{S}}D(P_1(\cdot|i)||P_2(\cdot|i)),~(K-1)\sum\limits_{i\in\mathcal{S}}D(P_2(\cdot|i)||P_1(\cdot|i))\bigg\rbrace,\label{eq:lower_bound_partial_10_rested_arms}
\end{align}\endgroup
where $(a)$ above follows by noting that $P_1'$ appears only in the term $D(P_2(\cdot|i)||P_1'(\cdot|i))$, and that for the choice $P_1'=P_2$, we get $D(P_2(\cdot|i)||P_1'(\cdot|i))=0$ for all $i\in\mathcal{S}$. For ease of notation, we shall denote the quantity in \eqref{eq:lower_bound_partial_10_rested_arms} by $\Delta'$, which we note is a constant.

Combining \eqref{eq:proof_of_prop_1_temp_1} with \eqref{eq:lower_bound_partial_9_rested_arms}, we get the following relation after rearrangement:
\begingroup\allowdisplaybreaks\begin{align}
	d(\epsilon,1-\epsilon)
	\leq \Delta' + (E^\pi[\tau+1|C]-K)\,\bigg\lbrace\max\limits_{\lambda\in\mathcal{P}(\mathcal{A})}\min\limits_{C'=(h',P_1',P_2')}\bigg[\sum\limits_{a=1}^{K}\lambda(a) D(P_h^a(\cdot|\cdot)||P_{h'}^a(\cdot|\cdot)|q_h^a)\bigg]\bigg\rbrace.\label{eq:proof_of_prop_1_temp_2}
\end{align}\endgroup

Since \eqref{eq:proof_of_prop_1_temp_2} is valid for any arbitrary choice of $\epsilon>0$ and for all $\pi\in\Pi(\epsilon)$, letting $\epsilon\downarrow 0$ and using $d(\epsilon,1-\epsilon)/\log \frac{1}{\epsilon} \to 1$ as $\epsilon\downarrow 0$, along with the fact that $q_h^a(i)\to \mu_h^a(i)$ for all $i\in\mathcal{S}$ in the regime of vanishing error probabilities, we get
 \begin{equation}
\lim\limits_{\epsilon\downarrow 0}\inf\limits_{\pi\in\Pi(\epsilon)}\frac{E^\pi[\tau(\pi)|C]}{\log\frac{1}{\epsilon}}\geq \frac{1}{D^*(h,P_1,P_2)},
\end{equation}
 	where the quantity $D^*(h,P_1,P_2)$ depends on the underlying configuration of the arms, and is given by
 	\begingroup\allowdisplaybreaks\begin{align}
 D^*(h,P_1,P_2)
=\max\limits_{\lambda\in\mathcal{P}(\mathcal{A})}\,\min\limits_{C'=(h',P_1',P_2')}\bigg(\sum\limits_{a=1}^{K}\lambda(a)\,D(P_h^a(\cdot|\cdot)||P_{h'}^a(\cdot|\cdot)|\mu_h^a)\bigg).\label{eq:D^*(h,P_1,P_2)}
 	\end{align}\endgroup
 	
We now show that the quantities in \eqref{eq:D^*(h,P_1,P_2)} and \eqref{eq:D^*(h,P_1,P_2)_simpl_final} are the same.
 	
\subsubsection{The Final Steps}
 Using \eqref{eq:P_h^a(j|i)} and \eqref{eq:mu_h^a}, and using the shorthand notation $D(P_h^a||P_{h'}^a|\mu_h^a)$ to denote the KL divergence term inside the summation in \eqref{eq:D^*(h,P_1,P_2)}, we get
  \begingroup\allowdisplaybreaks\begin{align}
 		&D^*(h,P_1,P_2)
 		&=\max\limits_{\lambda\in\mathcal{P}(\mathcal{A})}\,\min\limits_{h'\neq h,\,P_1',\,P_2'}\bigg(\lambda(h)\,D(P_1||P_2'|\mu_1)
 		&+\lambda(h')\,D(P_2||P_1'|\mu_2)+(1-\lambda(h)-\lambda(h'))D(P_2||P_2'|\mu_2)\bigg).\label{eq:D^*(h,P_1,P_2)_simpl_1}
 \end{align}\endgroup
 Since $P_1'$ appears only in the second term on right-hand side of the above expression, the minimum over all $P_1'$ of the quantity $D(P_2||P_1'|\mu_2)$ is equal to zero, which is attained for $P_1'=P_2$. Thus, we have

 \begingroup\allowdisplaybreaks\begin{align}
 		D^*(h,P_1,P_2)
 		=\max\limits_{\lambda\in\mathcal{P}(\mathcal{A})}\,\min\limits_{h'\neq h,\,P_2'}\bigg(\lambda(h)\,D(P_1||P_2'|\mu_1)
 		+(1-\lambda(h)-\lambda(h'))D(P_2||P_2'|\mu_2)\bigg).\label{eq:D^*(h,P_1,P_2)_simpl_2}
 \end{align}\endgroup
 We now note that
 \begingroup\allowdisplaybreaks\begin{align}
 	\min\limits_{h'\neq h}(1-\lambda(h)-\lambda(h'))&=1-\lambda(h)-\max\limits_{h'\neq h}\lambda(h')\nonumber\\
 	&\stackrel{(a)}{\leq} 1-\lambda(h)-\frac{1-\lambda(h)}{K-1}\nonumber\\
 	&=(1-\lambda(h))\frac{\left(K-2\right)}{(K-1)},\label{eq:min_over_h'_simplified}
 \end{align}\endgroup
 where $(a)$ above follows by lower bounding the maximum of a set of numbers by their arithmetic mean. We then have
 \begingroup\allowdisplaybreaks\begin{align}
 		D^*(h,P_1,P_2)
 		=\max\limits_{0\leq\lambda(h)\leq 1}\,\min\limits_{P_2'}\bigg(\lambda(h)\,D(P_1||P_2'|\mu_1)
 		+(1-\lambda(h))\frac{(K-2)}{(K-1)}D(P_2||P_2'|\mu_2)\bigg).\label{eq:D^*(h,P_1,P_2)_simpl_3}
 \end{align}\endgroup
 Using Lemma 3 in \eqref{eq:D^*(h,P_1,P_2)_simpl_3}, and recognising that the hand side of \eqref{eq:D^*(h,P_1,P_2)_simpl_3} is not a function of $h$, we write
 \begingroup\allowdisplaybreaks\begin{align}
 		D^*(h,P_1,P_2)
 		=\max\limits_{0\leq\lambda_1\leq 1}\bigg(\lambda_1\,D(P_1||P|\mu_1)+(1-\lambda_1)\frac{(K-2)}{(K-1)}D(P_2||P|\mu_2)\bigg),\label{eq:D^*(h,P_1,P_2)_simpl_final_4}
 \end{align}\endgroup
 where $P$ is a probability transition matrix whose entry in the $i$th row and $j$th column is given by
 \begin{equation}
 	P(j|i)=\frac{\lambda_1\mu_1(i)P_1(j|i)+(1-\lambda_1)\frac{(K-2)}{(K-1)}\mu_2(i)P_2(j|i)}{\lambda_1\mu_1(i)+(1-\lambda_1)\frac{(K-2)}{(K-1)}\mu_2(i)}.\label{eq:P_matrix_entries}
 \end{equation}
Noting that the right-hand sides of \eqref{eq:D^*(h,P_1,P_2)_simpl_final_4} and \eqref{eq:D^*(h,P_1,P_2)_simpl_final} are identical, this completes the proof of the proposition.
\end{IEEEproof}

\subsection{Proof of Proposition \ref{prop:positive_drift_of_M_{hh'}(n)}}\label{appndx:proof_of_strictly_positive_drift_of_M_{hh'}(n)}
Let $C=(h,P_1,P_2)$ be the underlying configuration of the arms. We first show in the following lemma that under the non-stopping version of policy $\pi^{\star}(L,\delta)$, the maximum likelihood estimates $\hat{P}^n_{1,h}$ and $\hat{P}^n_{h,2}$ converge to their respective true values $P_1$ and $P_2$.
\begin{lemma}\label{lemma:convergence_of_ML_estimates}
	Let $C=(h,P_1,P_2)$ denote the underlying configuration of the arms. Then, under the non-stopping version of policy $\pi^{\star}(L,\delta)$, as $n\to\infty$, the following convergences hold a.s. for all $i,j\in\mathcal{S}$:
	\begingroup\allowdisplaybreaks\begin{align}
		\frac{N_a(n,i,j)}{N_a(n,i)}\longrightarrow\begin{cases}
			P_1(j|i),&a=h,\\
			P_2(j|i),&a\neq h,
		\end{cases},\quad
		\frac{\sum\limits_{a\neq h}N_a(n,i,j)}{\sum\limits_{a\neq h}N_a(n,i)}\longrightarrow P_2(j|i).\label{eq:convergence_of_ml_estimates}
	\end{align}\endgroup
	\qed
\end{lemma}

\begin{IEEEproof}
Fix $i,j\in\mathcal{S}$ and $a\in\mathcal{A}$. Let $S_a(n)$ denote the quantity
\begin{equation}
	S_a(n)=\sum\limits_{t=0}^{n-1}\left(1_{\{A_{t+1}=a\}}-P(A_{t+1}=a|A^t,\bar{X}^t)\right),\label{eq:S_a(n)}
\end{equation}	
where $P(A_{t+1}=a|A^t,\bar{X}^t)$ is given by
	\begingroup\allowdisplaybreaks\begin{align}
	P(A_{t+1}=a|A^t,\bar{X}^t)
	=\frac{\delta}{K}+(1-\delta)\,\lambda^*(h^*(t),\hat{P}^t_{h^*(t),1},\hat{P}^t_{h^*(t),2})(a).\label{eq:P(A_{n+1}=a|Z^n,A^n)}
\end{align}\endgroup

\textcolor{black}{Letting $d^a_{t+1}=1_{\{A_{t+1}=a\}}-P(A_{t+1}=a|A^t,\bar{X}^t)$, we note that $P(|d_{t+1}|\leq 2|A^t,\bar{X}^t)=1$ for all $t\geq 0$, implying that $\{d_{t}\}_{t\geq 0}$ is bounded uniformly a.s.. Since $\{d_{t+1}\}_{t\geq 0}$ is a martingale difference sequence, it follows from \cite[Th. 1.2A]{Victor1999} that for every $\epsilon>0$, there exists $c_\epsilon>0$ such that $P(\frac{S_a(n)}{n}>\epsilon)\leq e^{-nc_\epsilon}$. From this, it follows that $S_a(n)/n\to 0$ a.s..} This implies that the following is true a.s. for sufficiently large values of $n$:
\begingroup\allowdisplaybreaks\begin{align}
	\frac{\delta}{2K}<\frac{N_a(n)-1}{n}<1+\frac{\delta}{2K}.\label{eq:N_a(n)_lies_between_two_quantities}
\end{align}\endgroup
Thus, we have $\liminf\limits_{n\to\infty}\frac{N_a(n)}{n}>\frac{\delta}{2K}>0$ a.s.. By the ergodic theorem, it then follows that as $n\to\infty$, the following convergences hold a.s.:
\begingroup\allowdisplaybreaks\begin{align}
    \frac{N_a(n,i)}{N_a(n)}\longrightarrow \mu_h^a(i),\quad
	\frac{N_a(n,i,j)/N_a(n)}{N_a(n,i)/N_a(n)}\longrightarrow P_h^a(j|i)\label{eq:first_part_convergence};
\end{align}\endgroup
here, $\mu_h^a(i)$ and $P_h^a(j|i)$ are as defined in \eqref{eq:mu_h^a} and \eqref{eq:P_h^a(j|i)} respectively.
This establishes the convergence in the first line of \eqref{eq:convergence_of_ml_estimates} under the assumption that $C=(h,P_1,P_2)$ is the underlying configuration of the arms.

We then note that a.s.,
\begingroup\allowdisplaybreaks\begin{align}
	\frac{\sum\limits_{a\neq h}N_a(n,i,j)}{\sum\limits_{a\neq h}N_a(n,i)}&=\frac{\sum\limits_{a\neq h}\frac{N_a(n,i,j)}{N_h^a(n,i)}\frac{N_h^a(n,i)}{N_h^a(n)}\frac{N_h^a(n)}{n}}{\sum\limits_{a\neq h}\frac{N_a(n,i)}{N_h^a(n)}\frac{N_h^a(n)}{n}}\nonumber\\
	&\stackrel{n\to\infty}{\longrightarrow} P_2(j|i),
\end{align}\endgroup
where the convergence in the last line above follows from \eqref{eq:first_part_convergence} by noting that for $a\neq h$, when $C=(h,P_1,P_2)$ is the underlying configuration of the arms, $\mu_h^a(i)=\mu_2(i)$ and $P_h^a(j|i)=P_2(j|i)$. This establishes the convergence in the second line of \eqref{eq:convergence_of_ml_estimates}, thus completing the proof of the lemma.
\end{IEEEproof}

\begin{IEEEproof}[Proof of Proposition \ref{prop:positive_drift_of_M_{hh'}(n)}]
We now use Lemma \ref{lemma:convergence_of_ML_estimates} to show that \eqref{eq:positive_drift_of_M_{hh'}(n)} holds for any $h'\neq h$. Towards this, we show that the quantity on the right-hand side of \eqref{eq:M_{hh'}(n)} is strictly positive.

For any choice of $\epsilon'>0$, we have the following:
\begin{enumerate}
	\item Since $T_1$ is a constant that does not grow with $n$, we have
	\begin{equation}
		\lim\limits_{n\to\infty}\frac{T_1}{n}=0,\label{eq:liminf_t_1(n)/n_final}
	\end{equation}
	and therefore it follows that there exists a positive integer $M_1=M_1(\epsilon')$ such that $T_1/n\geq -\epsilon'$ for all $n\geq M_1$.
	\item From \eqref{eq:t_2(n)}, we have
    \begingroup\allowdisplaybreaks\begin{align}
    	\frac{T_2(n)}{n}=\frac{1}{n}\sum\limits_{i\in\mathcal{S}}\log B((N_h(n,i,j)+1)_{j\in\mathcal{S}}).\label{eq:liminf_t_2(n)/n_1}
    \end{align}\endgroup
    Fix $i\in\mathcal{S}$. Then, we have
    \begingroup\allowdisplaybreaks\begin{align}
    	\log B((N_h(n,i,j)+1)_{j\in\mathcal{S}})=\log E\left[\prod\limits_{j\in\mathcal{S}}X_{ij}^{N_h(n,i,j)}\right],\label{eq:liminf_t_2(n)/n_2}
    \end{align}\endgroup
    where the random vector $(X_{ij})_{j\in\mathcal{S}}$ follows Dirichlet distribution with parameters $\alpha_j=1$ for all $j\in\mathcal{S}$. We now write \eqref{eq:liminf_t_2(n)/n_2} as follows:
    \begingroup\allowdisplaybreaks\begin{align}
    	\frac{1}{N_h(n)}\log B((N_h(n,i,j)+1)_{j\in\mathcal{S}})
    	=\frac{1}{N_h(n)}\log E\left[\exp\left(N_h(n)\sum\limits_{j\in\mathcal{S}} \frac{N_h(n,i,j)}{N_h(n)}\log X_{ij}\right)\right].\label{eq:liminf_t_2(n)/n_3}
    \end{align}\endgroup
    When $C=(h,P_1,P_2)$ is the underlying configuration of the arms, from Lemma \ref{lemma:convergence_of_ML_estimates}, we have that $N_h(n,i,j)/N_h(n)$ converges a.s. as $n\to\infty$ to $\mu_1(i)P_1(j|i)$. Thus, there exists a positive integer $M_{21}=M_{21}(\epsilon')$ such that for all $n\geq M_{21}$, we have
    \begingroup\allowdisplaybreaks\begin{align}
    	\frac{1}{N_h(n)}\log B((N_h(n,i,j)+1)_{j\in\mathcal{S}})
    	\geq \frac{1}{N_h(n)}\log E\left[\exp\left(N_h(n)\sum\limits_{j\in\mathcal{S}} (\mu_1(i)P_1(j|i)+\epsilon')\log X_{ij}\right)\right].\label{eq:liminf_t_2(n)/n_4}
    \end{align}\endgroup
    Noting that $N_h(n)$ converges a.s. to $+\infty$ as $n\to\infty$, by Varadhan's integral lemma \cite[Theorem 4.3.1]{AmirDembo2009}, there exists a positive integer $M_{22}=M_{22}(\epsilon')$ such that for all $n\geq M_2=\max\{M_{21},M_{22}\}$, we have
    \begingroup\allowdisplaybreaks\begin{align}
    	\frac{1}{N_h(n)}\log B((N_h(n,i,j)+1)_{j\in\mathcal{S}})
    	&\stackrel{(a)}{\geq} \sup\limits_{\{z_j\geq 0,\sum\limits_{j\in\mathcal{S}}z_j=1\}}\sum\limits_{j\in\mathcal{S}}(\mu_1(i)P_1(j|i)+\epsilon')\log z_j-\frac{\epsilon'}{|\mathcal{S}|}\nonumber\\
    	&=\sum\limits_{j\in\mathcal{S}}(\mu_1(i)P_1(j|i)+\epsilon')\log\frac{\mu_1(i)P_1(j|i)+\epsilon'}{\mu_1(i)+\epsilon'|\mathcal{S}|}-\frac{\epsilon'}{|\mathcal{S}|},\label{eq:liminf_t_2(n)/n_5}
    \end{align}\endgroup
    where the supremum on the right-hand side of $(a)$ above is computed over all vectors $(z_j)_{j\in\mathcal{S}}$ such that $z_j\geq 0$ for all $j\in\mathcal{S}$, and $\sum\limits_{j\in\mathcal{S}}z_j=1$. Plugging \eqref{eq:liminf_t_2(n)/n_5} into \eqref{eq:liminf_t_2(n)/n_1}, we get
    \begingroup\allowdisplaybreaks\begin{align}
    	\frac{T_{2}(n)}{n}
    	\geq \frac{N_h(n)}{n}\bigg\lbrace\bigg[\sum\limits_{i\in\mathcal{S}}\sum\limits_{j\in\mathcal{S}}(\mu_1(i)P_1(j|i)+\epsilon')
    	\log\frac{\mu_1(i)P_1(j|i)+\epsilon'}{\mu_1(i)+\epsilon'|\mathcal{S}|}\bigg]-\epsilon'\bigg\rbrace\label{eq:liminf_t_2(n)/n_final}
    \end{align}\endgroup
    for all $n\geq M_2$.
    \item From \eqref{eq:t_3(n)}, we have
    \begingroup\allowdisplaybreaks\begin{align}
    	\frac{T_3(n)}{n}=\frac{1}{n}\sum\limits_{i\in\mathcal{S}}\log B\left(\left(\sum\limits_{a\neq h}N_a(n,i,j)+1\right)_{j\in\mathcal{S}}\right).\label{eq:liminf_t_3(n)/n_1}
    \end{align}\endgroup
    Using the same arguments as those used to simplify \eqref{eq:liminf_t_2(n)/n_1}, we obtain the following: there exists a positive integer $M_3=M_3(\epsilon')$ such that for all $n\geq M_3$, we have
    \begingroup\allowdisplaybreaks\begin{align}
    	\frac{T_{3}(n)}{n}
    	\geq \frac{\sum\limits_{a\neq h}N_a(n)}{n}\bigg\lbrace\bigg[\sum\limits_{i\in\mathcal{S}}\sum\limits_{j\in\mathcal{S}}(\mu_2(i)P_2(j|i)+\epsilon')
    	\log\frac{\mu_2(i)P_2(j|i)+\epsilon'}{\mu_2(i)+\epsilon'|\mathcal{S}|}\bigg]-\epsilon'\bigg\rbrace.\label{eq:liminf_t_3(n)/n_final}
    \end{align}\endgroup
    \item From \eqref{eq:t_4(n)}, we have
    \begingroup\allowdisplaybreaks\begin{align}
    	\frac{T_4(n)}{n}=-\frac{1}{n}\sum\limits_{i,j\in\mathcal{S}}N_{h'}(n,i,j)\log\frac{N_{h'}(n,i,j)}{N_{h'}(n,i)}\label{eq:liminf_t_4(n)/n_1}.
    \end{align}\endgroup
If $N_h(n,i)=0$ for some state $i\in\mathcal{S}$ (in which case it follows that $N_h(n,i,j)=0$ for all $j\in\mathcal{S}$), or if $N_h(n,i,j)=0$\footnote{This may be the case if, for instance, $P_2(j|i)=0$ for some pair of states $i,j\in\mathcal{S}$.} for some pair of states $i,j\in\mathcal{S}$, then the corresponding terms in the summation in \eqref{eq:liminf_t_4(n)/n_1} will be of the form $0\log \frac{0}{0}$ or $0\log 0$ respectively, which we treat as zero by convention. Thus, without loss of generality, we assume that $N_h(n,i,j)>0$ for all $i,j\in\mathcal{S}$.

    Noting that $h'\neq h$, when the underlying configuration is $C=(h,P_1,P_2)$, from Lemma \ref{lemma:convergence_of_ML_estimates}, we have the following almost sure convergences (as $n\to\infty$):
    \begingroup\allowdisplaybreaks\begin{align}
    	\frac{N_{h'}(n,i,j)}{n}&\to \mu_2(i)P_2(j|i),\nonumber\\
    	\frac{N_{h'}(n,i,j)}{N_{h'}(n,i)}&\to P_2(j|i).\label{eq:liminf_t_4(n)/n_2}
    \end{align}\endgroup
    Using these in \eqref{eq:liminf_t_4(n)/n_1}, we get that there exists a positive integer $M_4=M_4(\epsilon')$ such that for all $n\geq M_4$, we have
    \begingroup\allowdisplaybreaks\begin{align}
    	\frac{T_4(n)}{n}\geq \sum\limits_{i,j\in\mathcal{S}}(\mu_2(i)P_2(j|i)-\epsilon')\log\frac{1}{P_2(j|i)+\epsilon'}.\label{eq:liminf_t_4(n)/n_final}
    \end{align}\endgroup
    \item Lastly, we present a simplification of the term $T_5(n)/n$. From \eqref{eq:t_5(n)}, we have
    \begingroup\allowdisplaybreaks\begin{align}
    	\frac{T_5(n)}{n}=-\frac{1}{n}\sum\limits_{i,j\in\mathcal{S}}\sum\limits_{a\neq h'}N_{a}(n,i,j)\log\frac{\sum\limits_{a\neq h'}N_{a}(n,i,j)}{\sum\limits_{a\neq h'}N_{a}(n,i)}.\label{eq:liminf_t_5(n)/n_1}
    \end{align}\endgroup
    For each $n$ and each $i,j\in\mathcal{S}$, we define $P_n(j|i)$ as the following quantity:
    \begin{equation}
    	P_n(j|i)=\frac{\sum\limits_{a\neq h'}N_a(n,i,j)}{\sum\limits_{a\neq h'}N_a(n,i)}.\label{eq:P_n(j|i)}
    \end{equation}
    Note that $P_n=(P_n(j|i))_{i,j\in\mathcal{S}}$ constitutes a valid probability transition matrix. From Lemma \ref{lemma:convergence_of_ML_estimates}, under the underlying configuration $C=(h,P_1,P_2)$, we note the following almost convergences as $n\to\infty$:
    \begingroup\allowdisplaybreaks\begin{align}
    	\frac{\sum\limits_{a\neq h,h'}N_{a}(n,i,j)}{\sum\limits_{a\neq h,h'}N_{a}(n,i)}\stackrel{n\to\infty}{\longrightarrow} P_2(j|i),\quad
  	\frac{\sum\limits_{a\neq h,h'}N_{a}(n,i)}{\sum\limits_{a\neq h,h'}N_{a}(n)}\stackrel{n\to\infty}{\longrightarrow} \mu_2(i).
    \end{align}\endgroup
    The above convergences then imply that there exists a positive integer $M_5=M_5(\epsilon')$ such that for all $n\geq M_5$, we have
    \begingroup\allowdisplaybreaks\begin{align}
    	\frac{T_5(n)}{n}\geq \frac{N_h(n)}{n}\sum\limits_{i,j\in\mathcal{S}}(\mu_1(i)P_1(j|i)-\epsilon')\log\frac{1}{P_n(j|i)}
    	+\frac{\sum\limits_{a\neq h,h'}N_a(n)}{n}\sum\limits_{i,j\in\mathcal{S}}(\mu_2(i)P_2(j|i)-\epsilon')\log\frac{1}{P_n(j|i)}.\label{eq:liminf_t_5(n)/n_final}
    \end{align}\endgroup
\end{enumerate}

Combining the results in \eqref{eq:liminf_t_1(n)/n_final}, \eqref{eq:liminf_t_2(n)/n_final}, \eqref{eq:liminf_t_3(n)/n_final}, \eqref{eq:liminf_t_4(n)/n_final} and \eqref{eq:liminf_t_5(n)/n_final}, we get that for all $n\geq M(\epsilon')=\max\{M_1,\dots,M_5\}$, we have
\begingroup\allowdisplaybreaks\begin{align}
	\frac{M_{hh'}(n)}{n}\geq f_{n}(\epsilon'),\label{eq:liminf_M_{hh'}(n)/n_1}
\end{align}\endgroup
	%
where $f_n(\epsilon')$ denotes the sum of the terms of the right-hand sides of \eqref{eq:liminf_t_1(n)/n_final}, \eqref{eq:liminf_t_2(n)/n_final}, \eqref{eq:liminf_t_3(n)/n_final}, \eqref{eq:liminf_t_4(n)/n_final} and \eqref{eq:liminf_t_5(n)/n_final}.

We now define $f_n(0)$ as the following quantity:
\begingroup\allowdisplaybreaks\begin{align}
	f_n(0)\coloneqq\frac{N_h(n)}{n}D(P_1||P_n|\mu_1)+\frac{\sum\limits_{a\neq h,h'}N_a(n)}{n}D(P_2||P_n||\mu_2).\label{eq:f_n(0)}
\end{align}\endgroup
Then, by continuity, we have that for any choice of $\epsilon>0$, there exists $\epsilon'>0$ such that $f_n(\epsilon')>f_n(0)-\epsilon$ for all sufficiently large values of $n$. From \eqref{eq:liminf_M_{hh'}(n)/n_1}, this implies that
\begin{equation}
	\frac{M_{hh'}(n)}{n}> f_{n}(0)-\epsilon
\end{equation}
for all sufficiently large values of $n$, from which it follows that
\begin{equation}
	\liminf\limits_{n\to\infty}\left[\frac{M_{hh'}(n)}{n}-f_n(0)\right]\geq  -\epsilon.
\end{equation}
Since the above equation is true for an arbitrary choice of $\epsilon$, letting $\epsilon\downarrow 0$, we get
\begin{equation}
	\liminf\limits_{n\to\infty} \frac{M_{hh'}(n)}{n}-\limsup\limits_{n\to\infty}f_n(0)\geq  0,
\end{equation}
from which it follows that
\begingroup\allowdisplaybreaks\begin{align}
	\liminf\limits_{n\to\infty}\frac{M_{hh'}(n)}{n}&\geq  \limsup\limits_{n\to\infty}f_n(0)\nonumber\\
	&\geq \liminf\limits_{n\to\infty}f_n(0)\nonumber\\
	&\geq \liminf\limits_{n\to\infty}\bigg\lbrace\frac{N_h(n)}{n}D(P_1||P_n|\mu_1)+\frac{\sum\limits_{a\neq h,h'}N_a(n)}{n}D(P_2||P_n|\mu_2)\bigg\rbrace\nonumber\\
	&\geq \liminf\limits_{n\to\infty}\bigg\lbrace\frac{N_h(n)}{n}\,D(P_1||P_n|\mu_1)\bigg\rbrace+\liminf\limits_{n\to\infty}\bigg\lbrace\frac{\sum\limits_{a\neq{h,h'}}N_a(n)}{n}\,D(P_2||P_n|\mu_2)\bigg\rbrace
	\label{eq:liminf_M_{hh'}(n)/n_strictly_positive}
\end{align}\endgroup
We now claim that $\sup\limits_{n\geq 0}D(P_1||P_n|\mu_1)<\infty$ a.s.. Indeed, we note that
\begingroup\allowdisplaybreaks\begin{align}
	P_n(j|i)&=\frac{\sum\limits_{a\neq h'}N_a(n,i,j)}{\sum\limits_{a\neq h'}N_a(n,i)}\nonumber\\
	&\geq \frac{\sum\limits_{a\neq h'}N_a(n,i,j)}{n}\nonumber\\
	&\geq \left(\frac{N_h(n)}{n}\right)\left(\frac{N_h(n,i)}{N_h(n)}\right)\left(\frac{N_h(n,i,j)}{N_h(n,i)}\right)+\left(\frac{\sum\limits_{a\neq h,h'}N_a(n)}{n}\right)\left(\frac{\sum\limits_{a\neq h,h'}N_a(n,i)}{\sum\limits_{a\neq h,h'}N_a(n)}\right)\left(\frac{\sum\limits_{a\neq h,h'}N_a(n,i,j)}{\sum\limits_{a\neq h,h'}N_a(n,i)}\right)\nonumber\\
	&\stackrel{(a)}{\geq} \left(\frac{\delta}{2K}\right)\left(\frac{\mu_1(i)\,P_1(j|i)}{2}\right)+(K-2)\left(\frac{\delta}{2K}\right)\left(\frac{\mu_2(i)\,P_2(j|i)}{2}\right)\nonumber\\
	&\stackrel{(b)}{\geq}\left(\frac{\delta}{2K}\right)\left(\frac{\mu_1(i)P_1(j|i)+\mu_2(i)P_2(j|i)}{2}\right)\nonumber\\
	&\geq \left(\frac{\delta}{2K}\right)\left(\min\bigg\lbrace\min\limits_{i\in\mathcal{S}}\mu_1(i),\,\min\limits_{i\in\mathcal{S}}\mu_2(i)\bigg\rbrace\right)\left(\frac{P_1(j|i)+P_2(j|i)}{2}\right)~a.s.\label{eq:liminf_M_{hh'}(n)/n_strictly_positive_1}
\end{align}\endgroup
for all sufficiently large values of $n$, where $(a)$ follows from \eqref{eq:N_a(n)_lies_between_two_quantities} and Lemma \ref{lemma:convergence_of_ML_estimates}, and $(b)$ follows by using the fact that the number of arms $K\geq 3$. It then follows that
\begingroup\allowdisplaybreaks\begin{align} D(P_1 & ||P_n|\mu_1) \\
&=\sum\limits_{i\in\mathcal{S}}\mu_1(i)\sum\limits_{j\in\mathcal{S}}P_1(j|i)\log\frac{P_1(j|i)}{P_n(j|i)}\nonumber\\
	&\leq \sum\limits_{i,j\in\mathcal{S}}\mu_1(i)\,P_1(j|i)\log\frac{P_1(j|i)}{\frac{P_1(j|i)+P_2(j|i)}{2}}+\sum\limits_{i,j\in\mathcal{S}}\mu_1(i)P_1(j|i)\log P_1(j|i)+\log\frac{1}{\left(\frac{\delta}{2K}\right)\left(\min\bigg\lbrace\min\limits_{i\in\mathcal{S}}\mu_1(i),\,\min\limits_{i\in\mathcal{S}}\mu_2(i)\bigg\rbrace\right)}\nonumber\\
	&=D\left(P_1\bigg|\bigg|\frac{P_1+P_2}{2}\bigg\vert \mu_1\right)+\sum\limits_{i\in\mathcal{S}}\mu_1(i)(-H(P_1(\cdot|i))+\log\frac{1}{\left(\frac{\delta}{2K}\right)\left(\min\bigg\lbrace\min\limits_{i\in\mathcal{S}}\mu_1(i),\,\min\limits_{i\in\mathcal{S}}\mu_2(i)\bigg\rbrace\right)}\nonumber\\
	&<\infty~a.s..\label{eq:liminf_M_{hh'}(n)/n_strictly_positive_2}
\end{align}\endgroup
On similar lines, it can be shown that $D(P_2||P_n|\mu_1)$ is bounded uniformly a.s. for all $n\geq 0$. Using the uniform boundedness property just proved, we may express \eqref{eq:liminf_M_{hh'}(n)/n_strictly_positive} as
\begingroup\allowdisplaybreaks\begin{align}
	\liminf\limits_{n\to\infty}\frac{M_{hh'}(n)}{n}&\geq \bigg\lbrace\liminf\limits_{n\to\infty}\frac{N_h(n)}{n}\bigg\rbrace\bigg\lbrace \liminf\limits_{n\to\infty}D(P_1||P_n|\mu_1)\bigg\rbrace+\bigg\lbrace\liminf\limits_{n\to\infty}\frac{\sum\limits_{a\neq h,h'}N_a(n)}{n}\bigg\rbrace\bigg\lbrace \liminf\limits_{n\to\infty}D(P_2||P_n|\mu_2)\bigg\rbrace\nonumber\\
	&\geq \left(\frac{\delta}{2K}\right)\left(\liminf\limits_{n\to\infty}D(P_1||P_n|\mu_1)+(K-2)\,\liminf\limits_{n\to\infty}D(P_2||P_n|\mu_2)\right)~a.s.,\label{eq:liminf_M_{hh'}(n)/n_strictly_positive_3}
\end{align}\endgroup
where the last line follows from \eqref{eq:N_a(n)_lies_between_two_quantities}.

Finally, we show that the first limit infimum term in \eqref{eq:liminf_M_{hh'}(n)/n_strictly_positive_3} is strictly positive, and note that an exactly parallel argument may be used to show that the second limit infimum term is also strictly positive. Suppose that $\liminf\limits_{n\to\infty}D(P_1||P_n|\mu_1)=0$ a.s.. By the property that KL divergence is zero if and only if the argument probability distributions are identical, it follows that there exists a subsequence $(n_k)_{k\geq 1}$ such that $P_{n_k}(j|i)\to P_1(j|i)$ as $k\to\infty$ a.s. for all $i,j\in\mathcal{S}$. We now fix attention to this subsequence, and note that by the property that the sequences $(N_h(n_k)/n_k)_{k\geq 1}$ and $(\sum\limits_{a\neq h,h'}N_a(n_k)/n_k)_{k\geq 1}$ are bounded, there exists a further subsequence $(n_{k_l})_{l\geq 1}$ of $(n_k)_{k\geq 1}$ such that the aforementioned bounded sequences admit limits, say $\alpha$ and $\beta$ respectively. From Lemma \ref{lemma:convergence_of_ML_estimates}, we then have the following convergence a.s. as $l\to\infty$:
\begingroup\allowdisplaybreaks\begin{align}
	P_{n_{k_l}}(j|i)\to \frac{\alpha\,\mu_1(i)\,P_1(j|i)+\beta \,\mu_2(i)\,P_2(j|i)}{\alpha\,\mu_1(i)+\beta\,\mu_2(i)}.\label{eq:P_{n_{k_l}}_convergence}
\end{align}\endgroup
However, we note that the right-hand side of \eqref{eq:P_{n_{k_l}}_convergence} is not equal to $P_1(j|i)$ whenever $P_2(j|i)>0$, thus resulting in a contradiction. This completes the proof of the proposition.
\end{IEEEproof}

\subsection{Proof of Proposition \ref{prop:pi_{LRMB}(L,delta)_belongs_to_Pi(epsilon)}}\label{appndx:pi_{LRMB}(L,delta)_belongs_to_Pi(epsilon)}
 The policy $\pi^{\star}(L,\delta)$ commits error if one of the following events is true:
\begin{enumerate}
	\item The policy never stops in finite time.
	\item The policy stops in finite time and declares $h'\neq h$ as the true index of the odd arm.
\end{enumerate}
The event in item $1$ above has zero probability as a consequence of Proposition \ref{prop:positive_drift_of_M_{hh'}(n)}.
Thus, the probability of error of policy $\pi=\pi^{\star}(L,\delta)$, which we denote by $P^\pi_e$, may be evaluated as follows: suppose $C=(h,P_1,P_2)$ is the underlying configuration of the arms. Then,
\begingroup\allowdisplaybreaks\begin{align}
	P^\pi_e =P^\pi(I(\pi)\neq h|C)
	=P^\pi\bigg(\exists~ n\text{ and }~h'\neq h\text{ such that }
	I(\pi)=h'\text{ and } \tau(\pi)=n\bigg\vert C\bigg).\label{eq:P_e_partial_1}
\end{align}\endgroup
We now let
\begingroup\allowdisplaybreaks\begin{align}
	\mathcal{R}_{h'}(n)\coloneqq\{\omega:\tau(\pi)(\omega)=n,\,I(\pi)(\omega)=h'\}\label{eq:R_{h'}(n)}
\end{align}\endgroup
denote the set of all sample paths for which the policy stops at time $n$ and declares $h'$ as the true index of the odd arm. Clearly, the collection $\{\mathcal{R}_{h'}(n):h'\neq h,\,n\geq 0\}$ is a collection of mutually disjoint sets. Therefore, we have
\begingroup\allowdisplaybreaks\begin{align}
P^\pi_e &=P^\pi\left(\bigcup\limits_{h'\neq h}\,\bigcup\limits_{n=0}^{\infty}\mathcal{R}_{h'}(n)\bigg\vert C\right)\nonumber\\
&= \sum\limits_{h'\neq h}\sum\limits_{n=0}^{\infty}P^\pi(\tau(\pi)=n,I(\pi)=h'|C)\nonumber\\
&= \sum\limits_{h'\neq h}\sum\limits_{n=0}^{\infty}~\int\limits_{\mathcal{R}_{h'}(n)}\,dP^\pi(\omega|C)\nonumber\\
&\stackrel{(a)}{=}\sum\limits_{h'\neq h}\sum\limits_{n=0}^{\infty}~\int\limits_{\mathcal{R}_{h'}(n)}f(A^n(\omega),\bar{X}^n(\omega)|H_h)\,\bigg[\prod\limits_{t=0}^n P_h(A_t|A^{t-1},\bar{X}^{t-1})\bigg]\,d(A^n(\omega),\bar{X}^n(\omega))\nonumber\\
&\stackrel{(b)}{\leq} \sum\limits_{h'\neq h}\sum\limits_{n=0}^{\infty}~\int\limits_{\mathcal{R}_{h'}(n)}\hat{f}(A^n(\omega),\bar{X}^n(\omega)|H_h)\bigg[\prod\limits_{t=0}^n P_h(A_t|A^{t-1},\bar{X}^{t-1})\bigg]\,d(A^n(\omega),\bar{X}^n(\omega))\nonumber\\
&\stackrel{(c)}{=}\sum\limits_{h'\neq h}\sum\limits_{n=0}^{\infty}~\bigg\lbrace\int\limits_{\mathcal{R}_{h'}(n)}e^{-M_{h'h}(n)}~{f}(A^n(\omega),\bar{X}^n(\omega)|H_{h'})\bigg[\prod\limits_{t=0}^n P_{h'}(A_t|A^{t-1},\bar{X}^{t-1})\bigg] d(A^n(\omega),\bar{X}^n(\omega))\bigg\rbrace\nonumber\\
&\leq \sum\limits_{h'\neq h}\sum\limits_{n=0}^{\infty}~\bigg\lbrace\int\limits_{\mathcal{R}_{h'}(n)}\frac{1}{(K-1)L}~dP^\pi(\omega|C')\bigg\rbrace\nonumber\\
&=\sum\limits_{h'\neq h}\frac{1}{(K-1)L}~P^\pi\left(\bigcup\limits_{n=0}^{\infty}\mathcal{R}_{h'}(n)\bigg|C'\right){\leq}~ \frac{1}{L},
\end{align}\endgroup
where in $(a)$ above, $P_h(A_t|A^{t-1},\bar{X}^{t-1})$ denotes the probability of selecting arm $A_t$ at time $t$ when the index of the odd arm is $h$, with the convention that at time $t=0$, this term represents $P_h(A_0)$; $(b)$ above follows by the definition of $\hat{f}$ in \eqref{eq:ml_likelihod_under_hyp_h}, and $(c)$ follows by using the fact that the probability of selecting an arm at any time $t$, based on the history of past arm selections and observations, is independent of the odd arm index, and is thus the same when the arm indexed by either $h$ or $h'$ is the odd arm. Setting $L=1/\epsilon$ gives $P_e^{\pi}\leq \epsilon$, thus proving that $\pi=\pi^{\star}(L,\delta)\in\Pi(\epsilon)$. This completes the proof of the proposition.
 \qed

 \subsection{Proof of Proposition \ref{prop:lim_M_h(n)/n_correct_drift}}\label{appndx:proof_of_prop_lim_M_h(n)/n_correct_drift}
Before we present the proof of Proposition \ref{prop:lim_M_h(n)/n_correct_drift}, we show that the odd arm chosen by the non-stopping version of policy $\pi^{\star}(L,\delta)$ is indeed the correct one. Further, we show that the arm selection frequencies under the same policy converge to the respective optimal values given in \eqref{eq:lambda^*(h,P_1,P_2)}.
\begin{prop}
	Let $C=(h,P_1,P_2)$ denote the underlying configuration of the arms. Fix $L\geq 1$ and $\delta\in(0,1)$, and consider the non-stopping version of policy $\pi^{\star}(L,\delta)$. For any $h'\neq h$ and $i,j\in\mathcal{S}$, let $P_n(j|i)$ be defined as in \eqref{eq:P_n(j|i)}, Then, the following convergences hold a.s. as $n\to\infty$.
	\begingroup\allowdisplaybreaks\begin{align}
	&h^*(n)\to h,\label{eq:h^*_converges_to_h}\\
	&\lambda_{opt}(h^*(n),\hat{P}^n_{h^*(n),1},\hat{P}^n_{h^*(n),2})\to\lambda_{opt}(h,P_1,P_2),\label{eq:lambda^*_converges_to_true_lambda^*_values}\\
	&\frac{N_a(n)}{n}\to \lambda^*_\delta(h,P_1,P_2)(a)\text{ for all }a\in\mathcal{A},\label{eq:arm_frequencies_converge}\\
	&P_{n}(j|i)\to P_\delta(j|i)\text{ for all }i,j\in\mathcal{S},\label{eq:P_n_converges_to_P}
	\end{align}\endgroup
where for each $a\in\mathcal{A}$ and each $i,j\in\mathcal{S}$, the quantity $\lambda_{\delta}^*(h,P_1,P_2)(a)$ and the term $P_\delta(j|i)$ in \eqref{eq:P_n_converges_to_P} are as defined in the statement of Proposition \ref{prop:lim_M_h(n)/n_correct_drift}.
\qed
\end{prop}
\begin{IEEEproof}
We already established that \eqref{eq:limsup_M_{h'}(n)_less_than_0} holds for all sufficiently large $n$. This establishes \eqref{eq:h^*_converges_to_h}, which in turn implies that
\begingroup\allowdisplaybreaks\begin{align}
\lambda_{opt}(h^*(n),\hat{P}^n_{h^*(n),1},\hat{P}^n_{h^*(n),2})\to \lambda_{opt}(h,P_1,P_2),
\end{align}\endgroup
because of the convergence of the maximum likelihood estimates shown in \eqref{eq:convergence_of_ml_estimates}, and the fact that $\lambda^*(h,P,Q)$ is jointly continuous in the pair $(P,Q)$, a fact that follows from Berge's Maximum Theorem \cite{Ausubel1993}. This establishes \eqref{eq:lambda^*_converges_to_true_lambda^*_values}.

We now proceed to show \eqref{eq:arm_frequencies_converge}. Towards this, we observe that from \eqref{eq:P(A_{n+1}=a|sigma(A^n,X^n)} and the convergence in \eqref{eq:lambda^*_converges_to_true_lambda^*_values}, we have
\begingroup\allowdisplaybreaks\begin{align}
	P(A_{n+1}=a|A^n,\bar{X}^n)
	&=\frac{\delta}{K}+(1-\delta)\,\lambda_{opt}(h^*(n),\hat{P}^n_{h^*(n),1},\hat{P}^n_{h^*(n),2})(a)\nonumber\\
	&\to \frac{\delta}{K}+(1-\delta)\lambda_{opt}(h,P_1,P_2)(a).
\end{align}\endgroup
We revisit the quantity $S_a(n)$ defined in \eqref{eq:S_a(n)}, and use the fact that $\frac{S_a(n)}{n}\to 0$ a.s. as $n\to\infty$ to obtain
\begingroup\allowdisplaybreaks\begin{align}
	\frac{N_a(n)}{n}&\to \frac{1}{n}\sum\limits_{t=0}^{n-1}P(A_{t+1}=a|A^t,\bar{X}^t)\nonumber\\
	&\to \frac{\delta}{K}+(1-\delta)\lambda_{opt}(h,P_1,P_2)(a).
\end{align}\endgroup

This establishes \eqref{eq:arm_frequencies_converge}.

Defining
\begin{equation}
	\alpha_n\coloneqq\frac{N_h(n)}{n},\quad \beta_n\coloneqq\frac{\sum\limits_{a\neq h,h'}N_a(n)}{n},\label{eq:alpha_n_beta_n}
\end{equation}
we note that the convergence in \eqref{eq:arm_frequencies_converge} implies in particular that
\begingroup\allowdisplaybreaks\begin{align}
	\alpha_n &\to \lambda^*_\delta(h,P_1,P_2)(h)=\frac{\delta}{K}+(1-\delta)\lambda^*=\lambda_\delta^*,\nonumber\\
	\beta_n &\to (K-2)\left(\frac{\delta}{K}+(1-\delta)\frac{1-\lambda^*}{K-1}\right)\nonumber\\
	&=\frac{(K-2)}{(K-1)}(1-\lambda_\delta^*).\label{eq:alpha_n_beta_n_convergence}
\end{align}\endgroup
Taking limits as $n\to\infty$ on both sides of \eqref{eq:P_n(j|i)}, and using the above limits for $\alpha_n$ and $\beta_n$,  we get the convergence in \eqref{eq:P_n_converges_to_P}, hence completing the proof of the proposition.
\end{IEEEproof}

\begin{IEEEproof}[Proof of Proposition \ref{prop:lim_M_h(n)/n_correct_drift}]
We recall from \eqref{eq:liminf_M_{hh'}(n)/n_strictly_positive} and \eqref{eq:f_n(0)} that
\begingroup\allowdisplaybreaks\begin{align}
	\liminf_{n\to\infty}\frac{M_{hh'}(n)}{n}
	&\geq \liminf\limits_{n\to\infty}\alpha_n D(P_1||P_n|\mu_1)+\liminf\limits_{n\to\infty}\beta_n D(P_2||P_n|\mu_2)\nonumber\\
	&=\lambda_\delta^*D(P_1||P_\delta|\mu_1)
	+\frac{(K-2)}{(K-1)}(1-\lambda_\delta^*)D(P_2||P_\delta||\mu_2),\label{eq:liminf_M_{hh'}(n)_D_delta}
\end{align}\endgroup
where the terms $\alpha_n$ and $\beta_n$ are as given in \eqref{eq:alpha_n_beta_n}.
Using Varadhan's integral lemma \cite[Theorem 4.3.1]{AmirDembo2009} to write
\begingroup\allowdisplaybreaks\begin{align}
    	\limsup\limits_{n\to\infty}\frac{1}{n}\log B((N_h(n,i,j)+1)_{j\in\mathcal{S}})
    	&\leq \limsup\limits_{n\to\infty}\frac{N_h(n)}{n}\mu_1(i)\sup\limits_{\{z_j\geq 0,\,\sum\limits_{j\in\mathcal{S}}z_j=1\}}\sum\limits_{j\in\mathcal{S}}P_1(j|i)\log z_j\nonumber\\
    	&=\lim\limits_{n\to\infty}\frac{N_h(n)}{n}\mu_1(i)(-H(P_1(\cdot|i))),\label{eq:limsup_t_2(n)/n_3}
    \end{align}\endgroup
    and following similar steps leading to \eqref{eq:liminf_t_2(n)/n_final}, we obtain
    \begingroup\allowdisplaybreaks\begin{align}
    \limsup_{n\to\infty}\frac{M_{hh'}(n)}{n}
	&\leq \lim\limits_{n\to\infty}\alpha_n D(P_1||P_n|\mu_1)+\lim\limits_{n\to\infty}\beta_n D(P_2||P_n|\mu_2)\nonumber\\
	&=\lambda_\delta^*D(P_1||P_\delta|\mu_1)
	+\frac{(K-2)}{(K-1)}(1-\lambda_\delta^*)D(P_2||P_\delta||\mu_2).\label{eq:limsup_M_{hh'}(n)_D_delta}
    \end{align}\endgroup
Combining \eqref{eq:liminf_M_{hh'}(n)_D_delta} and \eqref{eq:limsup_M_{hh'}(n)_D_delta}, we get the desired result.
\end{IEEEproof}

\subsection{Proof of Proposition \ref{prop:upper_bound}}\label{appndx:proof_of_upper_bound}
This section is organised as follows. We first show in Lemma \ref{lemma:stopping_time_of_policy_goes_to_infinity} that the stopping time of policy $\pi^\star(L,\delta)$ goes to infinity as the error probability vanishes (or as $L\to\infty$). We then exploit this to show that under policy $\pi^\star(L,\delta)$, the modified GLR statistic has the correct drift (see Lemma \ref{lemma:M_h(N(pi))/N(pi)_has_almost_correct_drift}). That is, we build on the result of Proposition \ref{prop:positive_drift_of_M_{hh'}(n)} and obtain the explicit limit for the modified GLR statistic for the regime of vanishing error probability. We then use the result of Lemma \ref{lemma:M_h(N(pi))/N(pi)_has_almost_correct_drift} to show in Lemma \ref{lemma:almost_sure_upper_bound_for_policy_pi_star} that the stopping time of policy $\pi^*(L,\delta)$ satisfies an asymptotic almost sure upper bound that matches with the right-hand side of \eqref{eq:upper_bound}. Finally, we establish that for any fixed $\delta\in(0,1)$, the family $\{\tau(\pi^\star(L,\delta))/\log L:L\geq 1\}$ is uniformly integrable, and as an intermediate step towards this, we establish in Lemma \ref{lemma:exp_bound}  an exponential upper bound for a certain probability term. Combining the almost sure limit of Lemma \ref{lemma:almost_sure_upper_bound_for_policy_pi_star} along with the uniform integrability result then yields the desired upper bound in \eqref{eq:upper_bound}.

\begin{lemma}\label{lemma:stopping_time_of_policy_goes_to_infinity}
	Let $C=(h,P_1,P_2)$ denote the underlying configuration of the arms. Fix $\delta\in(0,1)$. Then, under policy $\pi^{\star}(L,\delta)$, we have
	\begin{equation}
		\liminf\limits_{L\to\infty}\tau(\pi^{\star}(L,\delta))=\infty\text{ a.s.}\label{eq:stopping_time_of_policy_goes_to_infinity}
	\end{equation}
    \qed
\end{lemma}
\begin{IEEEproof}
	Since policy $\pi=\pi^{\star}(L,\delta)$ selects each of the $K$ arms in the first $K$ slots, in order to prove the lemma, we note that it suffices to prove the following statement:
\begin{equation}
	\text{for each $m\geq K$,}\quad \lim\limits_{L\to\infty}P^\pi(\tau(\pi)\leq m|C)=0.
\end{equation}
Fix $m\geq K$, and note that
\begingroup\allowdisplaybreaks\begin{align}
	\limsup\limits_{L\to\infty}\,P^\pi(\tau(\pi)\leq m|C)
	&=\limsup\limits_{L\to\infty}\,P^\pi\bigg(\exists~K\leq n\leq m \text{ and }\tilde{h}\in\mathcal{A}
	\text{ such that }M_{\tilde{h}}(n)>\log((K-1)L)\bigg|C\bigg)\nonumber\\
	&\leq \limsup\limits_{L\to\infty}\sum\limits_{\tilde{h}\in\mathcal{A}}\sum\limits_{n=K}^{m}P^\pi(M_{\tilde{h}}(n)>\log((K-1)L)|C)\nonumber\\
	&\leq \limsup\limits_{L\to\infty}\frac{1}{\log((K-1)L)}\sum\limits_{\tilde{h}\in\mathcal{A}}\sum\limits_{n=K}^{m}E^\pi[M_{\tilde{h}}(n)|C],\label{eq:stop_time_goes_to_infty_1}
\end{align}\endgroup
where the first inequality above follows from the union bound, and the second inequality follows from Markov's inequality.

We now show that for each $m\in\{K,\ldots,n\}$, the expectation term inside the summation in \eqref{eq:stop_time_goes_to_infty_1} is finite. Towards this, we have
\begingroup\allowdisplaybreaks\begin{align}
	M_{\tilde{h}}(n)&=\log\left(\frac{f(A^n,\bar{X}^n|H_{\tilde{h}})}{\max\limits_{h'\neq \tilde{h}}\hat{f}(A^n,\bar{X}^n|H_{h'})}\right)\nonumber\\
	&\leq \log\left(\frac{\hat{f}(A^n,\bar{X}^n|H_{\tilde{h}})}{\hat{f}(A^n,\bar{X}^n|H_{h'})}\right)\text{ for all }h'\neq \tilde{h}.\label{eq:mod_glr_upper_bounded_by_glr}
\end{align}\endgroup
Fix an arbitrary $h'\neq \tilde{h}$. We recognise that the logarithmic term in \eqref{eq:mod_glr_upper_bounded_by_glr} is the classical GLR test statistic of hypothesis $H_{\tilde{h}}$ with respect to hypothesis $H_{h'}$, given by
\begingroup\allowdisplaybreaks\begin{align}
	\log\left(\frac{\hat{f}(A^n,\bar{X}^n|H_{\tilde{h}})}{\hat{f}(A^n,\bar{X}^n|H_{h'})}\right)=S_1(n)+S_2(n)+S_3(n)+S_4(n),\label{eq:stop_time_goes_to_infty_2}
\end{align}\endgroup
where the terms $S_1(n),\ldots,S_4(n)$ appearing in \eqref{eq:stop_time_goes_to_infty_2} are as below.
\begin{enumerate}
	\item The term $S_1(n)$ is given by
	\begin{equation}
		S_1(n)=\sum\limits_{i,j\in\mathcal{S}}N_{\tilde{h}}(n,i,j)\log\frac{N_{\tilde{h}}(n,i,j)}{N_{\tilde{h}}(n,i)}.\label{eq:S_1(n)}
	\end{equation}
	\item The term $S_2(n)$ is given by
    \begin{equation}
		S_2(n)=\sum\limits_{i,j\in\mathcal{S}}\sum\limits_{a\neq \tilde{h}}N_a(n,i,j)\log\frac{\sum\limits_{a\neq \tilde{h}}N_a(n,i,j)}{\sum\limits_{a\neq \tilde{h}}N_a(n,i)}.\label{eq:S_2(n)}
		\end{equation}
	\item The term $S_3(n)$ is given by
    \begin{equation}
		S_3(n)=-\sum\limits_{i,j\in\mathcal{S}}N_{h'}(n,i,j)\log\frac{N_{h'}(n,i,j)}{N_{h'}(n,i)}.\label{eq:S_3(n)}
	\end{equation}
	\item The term $S_4(n)$ is given by
    \begin{equation}
		S_4(n)=-\sum\limits_{i,j\in\mathcal{S}}\sum\limits_{a\neq h'}N_a(n,i,j)\log\frac{\sum\limits_{a\neq h'}N_a(n,i,j)}{\sum\limits_{a\neq h'}N_a(n,i)}.\label{eq:S_4(n)}
		\end{equation}
\end{enumerate}
We now obtain an a.s. upper bound for \eqref{eq:stop_time_goes_to_infty_2}. We recognise that $S_1(n)$ and $S_2(n)$ are non-positive, and thus upper bound each of these terms by zero. Let $$A(i)=(N_{h'}(n,i,j)/N_{h'}(n,i))_{j\in\mathcal{S}}$$ denote the probability vector corresponding to state $i$. Then, denoting the Shannon entropy of $A(i)$ by $H(A(i))$, we may express $S_3(n)$ as
\begingroup\allowdisplaybreaks\begin{align}
	S_3(n)&=(N_{h'}(n)-1)\sum\limits_{i\in\mathcal{S}}\bigg[\frac{N_{h'}(n,i)}{N_{h'}(n)-1}\bigg]H(A(i))\nonumber\\
	&\leq (N_{h'}(n)-1)~H\left(\sum\limits_{i\in\mathcal{S}}\bigg[\frac{N_{h'}(n,i)}{N_{h'}(n)-1}\bigg]A(i)\right)\nonumber\\
	&\leq N_{h'}(n) \log|\mathcal{S}|,\label{eq:S_3(n)_upper_bound}
\end{align}\endgroup
where the first inequality above follows from the concavity of the entropy function $H(\cdot)$, and the second inequality follows by noting that the Shannon entropy of a probability distribution on an alphabet of size $R$ is upper bounded by $\log R$. On similar lines, we get
\begingroup\allowdisplaybreaks\begin{align}
	S_4(n)&\leq \left(\sum\limits_{a\neq h'}N_a(n)\right)\log|\mathcal{S}|.\label{eq:S_4(n)_upper_bound}
\end{align}\endgroup
Using in \eqref{eq:stop_time_goes_to_infty_2} the results of \eqref{eq:S_3(n)_upper_bound} and \eqref{eq:S_4(n)_upper_bound}, along with the zero upper bound for the non-positive terms in \eqref{eq:S_1(n)} and \eqref{eq:S_2(n)} and the relation \eqref{eq:sum_N_a}, we get
\begingroup\allowdisplaybreaks\begin{align}
	M_{\tilde{h}}(n)\leq (n+1)\log |\mathcal{S}|\text{ a.s.},\label{eq:upper_bound_on_M_h(n)}
\end{align}\endgroup
from which it follows that
\begingroup\allowdisplaybreaks\begin{align}
	\limsup\limits_{L\to\infty}P^\pi(\tau(\pi)\leq m|C)
	&\leq \limsup\limits_{L\to\infty}\frac{1}{\log((K-1)L)}\sum\limits_{\tilde{h}\in\mathcal{A}}\sum\limits_{n=K}^{m}(n+1)\log |\mathcal{S}|\nonumber\\
	&=0.
\end{align}\endgroup
This completes the proof of the lemma.
\end{IEEEproof}

\begin{lemma}\label{lemma:M_h(N(pi))/N(pi)_has_almost_correct_drift}
	Let $C=(h,P_1,P_2)$ denote the underlying configuration of the arms. Fix $\delta\in(0,1)$. Then, under policy $\pi=\pi^{\star}(L,\delta)$, for any $h'\neq h$, we have
	\begin{equation}
		\lim\limits_{L\to\infty}\frac{M_{hh'}(\tau(\pi))}{\tau(\pi)}=D_\delta^*(h,P_1,P_2)~~a.s.\label{eq:M_h(N(pi))/N(pi)_has_almost_correct_drift}
	\end{equation}
	\qed
\end{lemma}
\begin{IEEEproof}
The proof follows as a consequence of Proposition \ref{prop:lim_M_h(n)/n_correct_drift} and Lemma \ref{lemma:stopping_time_of_policy_goes_to_infinity}.
\end{IEEEproof}

\begin{lemma}\label{lemma:almost_sure_upper_bound_for_policy_pi_star}

Let $C=(h,P_1,P_2)$ denote the underlying configuration of the arms. Fix $\delta\in(0,1)$. Then, under policy $\pi=\pi^*(L,\delta)$, we have
\begingroup\allowdisplaybreaks\begin{align}
	\limsup\limits_{L\to\infty}\, \frac{\tau(\pi)}{\log L}\leq  \frac{1}{D_\delta^*(h,P_1,P_2)}\quad a.s.\label{eq:almost_sure_upper_bound_for_policy_pi_star}
\end{align}\endgroup	
\end{lemma}
\begin{IEEEproof}
We first show that for any $h'\neq h$ and $n\geq 1$, the increment $M_{hh'}(n)-M_{hh'}(n-1)$ is bounded. Fix an arbitrary $h'\neq h$, and consider the following cases.
	\begin{enumerate}
		\item Case 1: Suppose that arm $h$ is selected at time $n$. Then, noting that in the expression for $M_{hh'}(n)$, the only terms that depend on the arm index $h$ are those in \eqref{eq:t_2(n)} and \eqref{eq:t_5(n)}, we have
		    \begin{equation}
		    	M_{hh'}(n)-M_{hh'}(n-1)=\bigg[T_2(n)-T_2(n-1)\bigg] + \bigg[T_5(n)-T_5(n-1)\bigg].\label{eq:mod_glr_bdd_incr_1}
		    \end{equation}
		    Suppose that at time $n$, the Markov process of arm $h$ undergoes a transition from state $i$ to state $j$, where $i,j\in\mathcal{S}$ are such that $\max\{P_1(j|i),P_2(j|i)\}>0$\footnote{Otherwise, a jump from $i$ to $j$ is not observed on arm $h$.}. Then, noting that
		    \begingroup\allowdisplaybreaks\begin{align}
		    	N_a(n,i',j')&=N_a(n-1,i',j')\quad \text{for all }a\in\mathcal{A},~i'\neq i,~j'\neq j,\nonumber\\
		    	N_h(n,i,j)&=N_h(n-1,i,j)+1,\nonumber\\
		    	N_a(n,i')& =N_a(n-1,i')\quad \text{for all }a\in\mathcal{A},~i'\neq i,\nonumber\\
		    	N_h(n,i)&=N_h(n-1,i)+1,
		    \end{align}\endgroup
		    it can be shown after some simplification that
		    \begingroup\allowdisplaybreaks\begin{align}
		    	T_2(n)-T_2(n-1)&=\log\frac{B(N_h(n-1,i,j)+2,(N_h(n-1,i,j')+1)_{j'\neq j})}{B(N_h(n-1,i,j')+1)_{j'\in\mathcal{S}}}\nonumber\\
		    	&\stackrel{(a)}{=}\frac{N_h(n-1,i,j)}{\sum\limits_{j'\in\mathcal{S}}N_h(n-1,i,j')}\nonumber\\
		    	&\leq 1\quad a.s.,
		    \end{align}\endgroup
		    where $(a)$ above follows by using the relation
		    \begin{equation}
		    	B(\alpha_1,\ldots,\alpha_{|\mathcal{S}|})=\left({\prod\limits_{k=1}^{|\mathcal{S}|}\Gamma(\alpha_k)}\right)\bigg/{\Gamma\left(\sum\limits_{k=1}^{|\mathcal{S}|}\alpha_k\right)}.
		    \end{equation}
		    Also, we have
		    \begingroup\allowdisplaybreaks\begin{align}
		    	T_5(n)-T_5(n-1)&=\left(\sum\limits_{a\neq h'}N_a(n-1,i,j)\right)\log\frac{\sum\limits_{a\neq h'}N_a(n-1,i,j)}{\sum\limits_{a\neq h'}N_a(n-1,i)}\nonumber\\
		    	&\hspace{3cm}-\left(1+\sum\limits_{a\neq h'}N_a(n-1,i,j)\right)\log \frac{1+\sum\limits_{a\neq h'}N_a(n-1,i,j)}{1+\sum\limits_{a\neq h'}N_a(n-1,i)}\nonumber\\
		    	&\leq \log\frac{\sum\limits_{a\neq h'}N_a(n-1,i)}{\sum\limits_{a\neq h'}N_a(n,i,j)}\nonumber\\
		    	&\to\log \frac{1}{P_\delta(j|i)}\quad a.s.,
		    \end{align}\endgroup
		    where the convergence in the last line follows from \eqref{eq:P_n_converges_to_P}. Thus, it follows that the increment $M_{hh'}(n)-M_{hh'}(n-1)$ is bounded for all $n\geq 1$.
		    \item Case 2: Suppose that arm $h'$ is sampled at time $n$. Noting that the only terms that depend on the arm index $h'$ are those in \eqref{eq:t_3(n)} and \eqref{eq:t_4(n)},
		      the analysis for this case proceeds on the exactly same lines as that of Case 1 presented above, and is omitted.
		    \item Case 3: Suppose that arm $a'$ is sampled at time $n$, where $a'\in\mathcal{A}\setminus\{h,h'\}$. Noting that the only terms that depend on the arm index $a'$ are those in \eqref{eq:t_3(n)} and \eqref{eq:t_5(n)},
		      the analysis for this case proceeds on the exactly same lines as that of Case 1 presented above, and is omitted.
	\end{enumerate}
This establishes that the increments of the modified GLR process are bounded at all times.

Fix an arbitrary $h'\neq h$. By the definition of stopping time $\tau(\pi)$, we have that $M_{hh'}(\tau(\pi)-1)<\log ((K-1)L)$. Using this, we have
\begingroup\allowdisplaybreaks\begin{align}
	\limsup\limits_{L\to\infty}\frac{M_{hh'}(\tau(\pi))}{\log L}&\stackrel{(a)}{=}\limsup\limits_{L\to\infty}\frac{M_{hh'}(\tau(\pi)-1)}{\log L}\nonumber\\
	&\leq \limsup\limits_{L\to\infty}\frac{\log((K-1)L)}{\log L}\nonumber\\
	&=1\quad a.s.,\label{eq:limsup_M_hh'(tau)/log_L}
\end{align}\endgroup	
where $(a)$ above is due to boundedness of the increments of the modified GLR process established above. Then, using Lemma \ref{lemma:M_h(N(pi))/N(pi)_has_almost_correct_drift} along with the relation \eqref{eq:limsup_M_hh'(tau)/log_L} yields
\begingroup\allowdisplaybreaks\begin{align}
	\limsup\limits_{L\to\infty}\frac{\tau(\pi)}{\log L}&=\limsup\limits_{L\to\infty}\bigg\lbrace\left(\frac{\tau(\pi)}{M_{hh'}(\tau(\pi))}\right)\left(\frac{M_{hh'}(\tau(\pi))}{\log L}\right)\bigg\rbrace \nonumber\\
	&=\left(\lim\limits_{L\to\infty}\frac{\tau(\pi)}{M_{hh'}(\tau(\pi))}\right)\left(\limsup\limits_{L\to\infty}\frac{M_{hh'}(\tau(\pi))}{\log L}\right)\nonumber\\
	&\leq \frac{1}{D_\delta^*(h,P_1,P_2)}\quad a.s.,
\end{align}\endgroup
thus completing the proof of the lemma.
\end{IEEEproof}

\begin{IEEEproof}[Proof of Proposition \ref{prop:upper_bound}]
For any fixed $\delta\in(0,1)$, we now establish that under policy $\pi=\pi^\star(L,\delta)$, the family $\{\tau(\pi)/\log L:L\geq 1\}$ is uniformly integrable. In order to do so, we note that it suffices to show that
\begin{equation}
	\limsup\limits_{L\to\infty}E^\pi\bigg[\exp\bigg(\frac{\tau(\pi)}{\log L}\bigg)\bigg|C\bigg]<\infty.
\end{equation}
Towards this, let $l(L,\delta)$ denote the quantity
\begingroup\allowdisplaybreaks\begin{align}
	l(L,\delta)\coloneqq\frac{3\log((K-1)L)}{\frac{\delta}{2K}\bigg(D(P_1||P_\delta|\mu_1)+D(P_2||P_\delta|\mu_2)\bigg)}.\label{eq:l(L,delta)}
\end{align}\endgroup
Let $C=(h,P_1,P_2)$ be the underlying configuration of the arms. Further, let $\pi^\star_h=\pi^\star_h(L,\delta)$ denote the version of policy $\pi^{\star}(L,\delta)$ that stops only upon declaring $h$ as the index of the odd arm. Let
\begin{equation}
	u(L)\coloneqq\exp\bigg(\frac{1+l(L,\delta)}{\log L}\bigg)
\end{equation}
Clearly, we have $\tau(\pi^\star_h)\geq \tau(\pi)$ a.s.. Then,
\begingroup\allowdisplaybreaks\begin{align}
	\limsup\limits_{L\to\infty}E^\pi\bigg[\exp\bigg(\frac{\tau(\pi)}{\log L}\bigg)\bigg|C\bigg]
	&=\limsup\limits_{L\to\infty}\int\limits_{0}^{\infty}P^\pi\bigg(\frac{\tau(\pi)}{\log L}>\log x\bigg|C\bigg)\,dx\nonumber\\
	&\leq \limsup\limits_{L\to\infty}\int\limits_{0}^{\infty}P^\pi\bigg({\tau(\pi^\star_h)}\geq \lceil(\log x)({\log L})\rceil\bigg|C\bigg)\,dx\nonumber\\
	&\stackrel{(a)}{\leq} \limsup\limits_{L\to\infty}\bigg\lbrace u(L)+\int\limits_{u(L)}^{\infty}P^\pi\bigg({\tau(\pi^\star_h)}\geq \lceil(\log x)({\log L})\rceil\bigg|C\bigg)\,dx\bigg\rbrace\nonumber\\
	&\leq \exp\bigg(\frac{3}{\frac{\delta}{2K}(D(P_1||P_\delta|\mu_1)+D(P_2||P_\delta|\mu_2))}\bigg)\nonumber\\
	&\hspace{2cm}+\limsup\limits_{L\to\infty}\sum\limits_{n\geq l(L,\delta)}\exp\bigg(\frac{n+1}{\log L}\bigg)\,P^\pi(M_h(n)<\log((K-1)L)|C),\label{eq:uniform_integrability_1}
\end{align}\endgroup
where $(a)$ above follows by upper bounding the probability term by $1$ for all $x\leq u(L)$.

We now show that for all $n\geq l(L,\delta)$, the probability term in \eqref{eq:uniform_integrability_1} decays exponentially in $n$. This is a strengthening of the result in Proposition \ref{prop:positive_drift_of_M_{hh'}(n)} which only establishes that when $C=(h,P_1,P_2)$ is the underlying configuration of the arms, $M_h(n)\to\infty$ as $n\to\infty$.

\begin{lemma}\label{lemma:exp_bound}
	Let $C=(h,P_1,P_2)$ denote the underlying configuration of the arms. Fix $L\geq 1$, $\delta\in(0,1)$, and consider the policy $\pi=\pi^{\star}(L,\delta)$. There exist constants $\theta>0$ and $0<B<\infty$ independent of $L$ such that for all sufficiently large values of $n$, we have
	\begin{equation}
		P^\pi(M_h(n)<\log((K-1)L)|C)\leq Be^{-\theta n}.\label{eq:exponential_bound}
	\end{equation}
	\qed
\end{lemma}
\begin{IEEEproof}
Since
\begingroup\allowdisplaybreaks\begin{align}
	P^\pi(M_h(n)<\log((K-1)L)|C)
	&=P^\pi\left(\min\limits_{h'\neq h}M_{hh'}(n)<\log((K-1)L)\bigg|C\right)\nonumber\\
	&\leq \sum\limits_{h'\neq h}P^\pi\left(M_{hh'}(n)<\log((K-1)L)\bigg|C\right),\label{eq:exp_bound_1}
\end{align}\endgroup
in order to prove the lemma, it suffices to show that each term inside the summation in \eqref{eq:exp_bound_1} is exponentially bounded. Going further, we drop the superscript $\pi$ and the conditioning on configuration $C$ in $P^\pi(\cdot|C)$ for ease of notation. For all $i,j\in\mathcal{S}$, let
\begingroup\allowdisplaybreaks\begin{align}
	\tilde{P}_{n}(j|i)\coloneqq \frac{\alpha_n \mu_1(i) P_1(j|i)+\beta_n \mu_2(i)P_2(j|i)}{\alpha_n \mu_1(i)+\beta_n \mu_2(i)},\label{eq:tilde_P_n}
\end{align}\endgroup
where $\alpha_n$ and $\beta_n$ are as in \eqref{eq:alpha_n_beta_n}. Fix $h'\neq h$ and $\epsilon>0$ arbitrarily. Then, using \eqref{eq:M_{hh'}(n)} and triangle inequality, we have
\begingroup\allowdisplaybreaks\begin{align}
	P(M_{hh'}(n)<\log((K-1)L))
	&\leq U_1+U_2+U_3+U_4+U_5+U_6+U_7,\label{eq:P(M_{hh'}(n)<log((K-1)L))}
\end{align}\endgroup
where the terms $U_1,\ldots,U_7$ in \eqref{eq:P(M_{hh'}(n)<log((K-1)L))} are as below.
\begin{enumerate}
	\item The term $U_1$ is given by
	\begin{equation}
		U_1=P\left(\frac{T_1(n)}{n}<-\epsilon\right)\label{eq:exp_bound_t_1/n},
	\end{equation}
	where $T_1$ is given by \eqref{eq:t_1}.
	\item The term $U_2$ is given by
	\begin{equation}
		U_2=P\left(\frac{T_2(n)}{n}-\frac{N_h(n)}{n}\sum\limits_{i\in\mathcal{S}}\mu_1(i)(-H(P_1(\cdot|i)))<-\epsilon\right)\label{eq:exp_bound_t_2(n)/n},
	\end{equation}
	where $T_2(n)$ is given by \eqref{eq:t_2(n)}.
    \item The term $U_3$ is given by
    \begin{equation}
    	U_3=P\left(\frac{T_3(n)}{n}-\frac{\sum\limits_{a\neq h}N_a(n)}{n}\sum\limits_{i\in\mathcal{S}}\mu_2(i)(-H(P_2(\cdot|i)))<-\epsilon\right)\label{eq:exp_bound_t_3(n)/n},
    \end{equation}
    where $T_3(n)$ is given by \eqref{eq:t_3(n)}.
    \item The term $U_4$ is given by
    \begin{equation}
    	U_4=P\left(\frac{T_4(n)}{n}-\frac{N_{h'}(n)}{n}\sum\limits_{i\in\mathcal{S}}\mu_2(i)H(P_2(\cdot|i))<-\epsilon\right)\label{eq:exp_bound_t_4(n)/n},
    \end{equation}
    where $T_4(n)$ is given by \eqref{eq:t_4(n)}.
    \item The term $U_5$ is given by
    \begin{equation}
    	U_5=P\left(\frac{T_5(n)}{n}-\sum\limits_{i\in\mathcal{S}}(\alpha_n\mu_1(i)+\beta_n\mu_2(i))H(\tilde{P}_n(\cdot|i))<-\epsilon\right)\label{eq:exp_bound_t_5(n)/n},
    \end{equation}
    where $T_5(n)$ is given by \eqref{eq:t_5(n)}.
    \item The term $U_6$ is given by
    \begin{equation}
    	U_6=P\bigg(\alpha_n \bigg[D(P_1||\tilde{P}_n|\mu_1)-D(P_1||P_\delta|\mu_1)\bigg]+\beta_n \bigg[D(P_2||\tilde{P}_n|\mu_2)-D(P_2||P_\delta|\mu_2)\bigg]<-\epsilon\bigg),\label{eq:exp_bound_t_6}
    \end{equation}
    where $P_\delta$ is the probability transition matrix described in the statement of Proposition \ref{prop:lim_M_h(n)/n_correct_drift}.
    \item The term $U_7$ is given by
    \begin{equation}
    	U_7=P\bigg(\alpha_n D(P_1||P_\delta|\mu_1)+\beta_n D(P_2||P_\delta|\mu_2)-6\epsilon
	<\frac{\log((K-1)L)}{n}\bigg).\label{eq:exp_bound_compensatory_term}
    \end{equation}
\end{enumerate}
In \eqref{eq:exp_bound_t_2(n)/n}, the term $H(P_1(\cdot|i))$ refers to the Shannon entropy of the probability distribution $(P_1(j|i))_{j\in\mathcal{S}}$ on set $\mathcal{S}$;
the terms $H(P_2(\cdot|i))$ and $H(\tilde{P}_n(\cdot|i))$ are defined similarly.

We now obtain a bound for the terms in \eqref{eq:exp_bound_t_1/n}-\eqref{eq:exp_bound_compensatory_term}.

\begin{enumerate}
\item We begin by showing an exponential upper bound for \eqref{eq:exp_bound_compensatory_term}. We choose $0<\epsilon'<\frac{2}{3}$, and then select $\epsilon>0$ such that the following holds:
\begingroup\allowdisplaybreaks\begin{align}
	\frac{\delta}{2K}(1-\epsilon')\bigg(D(P_1||P_\delta|\mu_1)+D(P_2||P_\delta|\mu_2)\bigg)-6\epsilon
	>\frac{1}{3}\cdot\frac{\delta}{2K}\bigg(D(P_1||P_\delta|\mu_1)+D(P_2||P_\delta|\mu_2)\bigg).\label{eq:select_epsilon'}
\end{align}\endgroup
Then, for all $n\geq l(L,\delta)$,
we have
\begingroup\allowdisplaybreaks\begin{align}
	P\bigg(\alpha_n D(P_1||P_\delta|\mu_1)+\beta_n D(P_2||P_\delta|\mu_2)-6\epsilon
	<\frac{\log((K-1)L)}{n},~
	\frac{N_a(n)}{n}>\frac{\delta}{2K}(1-\epsilon')\text{ for all }a\in\mathcal{A}\bigg)=0.\label{eq:exp_bound_compensatory_term_first_part_equal_zero}
\end{align}\endgroup
Writing the probability term in \eqref{eq:exp_bound_compensatory_term} as a sum of the probability term in \eqref{eq:exp_bound_compensatory_term_first_part_equal_zero} and a second probability term given by
\begin{equation}
	P\bigg(\alpha_n D(P_1||P_\delta|\mu_1)+\beta_n D(P_2||P_\delta|\mu_2)-6\epsilon
	<\frac{\log((K-1)L)}{n},~
	\frac{N_a(n)}{n}\leq\frac{\delta}{2K}(1-\epsilon')\text{ for some }a\in\mathcal{A}\bigg),\label{eq:exp_bound_compensatory_term_second_part}
\end{equation}
and upper bounding \eqref{eq:exp_bound_compensatory_term_second_part} by $P(N_a(n)/n~\leq (\delta/2K)(1-\epsilon')\text{ for some }a\in\mathcal{A})$, an application of the union bound yields
\begingroup\allowdisplaybreaks\begin{align}
	P\bigg(\alpha_n D(P_1||P_\delta|\mu_1)+\beta_n D(P_2||P_\delta|\mu_2)-6\epsilon
	<\frac{\log((K-1)L)}{n}\bigg)
	&\leq \sum\limits_{a=1}^{K} P\left(\frac{N_a(n)}{n}\leq\frac{\delta}{2K}(1-\epsilon')\right).\label{eq:exp_bound_correction_term_temp_1}
\end{align}\endgroup	
Noting that for each $a\in\mathcal{A}$, the sequence $\left(N_a(n)-n\frac{\delta}{2K}\right)_{n\geq 0}$ is a submartingale, with the absolute value of the difference between any two successive terms of the submartingale sequence being of value at most $1$, we use the Azuma-Hoeffding inequality to obtain
\begingroup\allowdisplaybreaks\begin{align}
	P\left(\frac{N_a(n)}{n}\leq\frac{\delta}{2K}(1-\epsilon')\right)&=P\bigg(N_a(n)-n\frac{\delta}{2K}\leq -n\epsilon'\frac{\delta}{2K}\bigg)\nonumber\\
	&= P\bigg(\bigg[N_a(n)-n\frac{\delta}{2K}\bigg]-N_a(0)\leq -n\epsilon'\frac{\delta}{2K}-N_a(0)\bigg)\nonumber\\
	& \leq P\bigg(\bigg[N_a(n)-n\frac{\delta}{2K}\bigg]-N_a(0)\leq -n\epsilon'\frac{\delta}{2K}\bigg)\nonumber\\
	&\leq \exp\left(-\frac{n(\epsilon')^2\delta^2}{8K^2}\right).\label{eq:azuma_bound_exp_bound_correction_term}
\end{align}\endgroup
Plugging \eqref{eq:azuma_bound_exp_bound_correction_term} back in \eqref{eq:exp_bound_correction_term_temp_1}, we arrive at
\begingroup\allowdisplaybreaks\begin{align}
	P\bigg(\alpha_n D(P_1||P_\delta|\mu_1)+\beta_n D(P_2||P_\delta|\mu_2)-6\epsilon
	<\frac{\log((K-1)L)}{n}\bigg)
	\leq K\exp\left(-\frac{n(\epsilon')^2\delta^2}{8K^2}\right).
\end{align}\endgroup

\item We now turn attention to \eqref{eq:exp_bound_t_4(n)/n}, which we upper bound as follows:
\begingroup\allowdisplaybreaks\begin{align}
	&P\left(\frac{T_4(n)}{n}-\frac{N_{h'}(n)}{n}\sum\limits_{i\in\mathcal{S}}\mu_2(i)H(P_2(\cdot|i))<-\epsilon\right)\nonumber\\
	&=P\bigg(\frac{N_{h'}(n)}{n}\bigg\lbrace\sum\limits_{i\in\mathcal{S}}\frac{N_{h'}(n,i)}{N_{h'}(n)}H\left(\frac{N_{h'}(n,i,\cdot)}{N_{h'}(n,i)}\right)
	-\mu_2(i)H(P_2(\cdot|i))\bigg\rbrace<-\epsilon\bigg)\nonumber\\
	&\leq P\bigg(\frac{N_{h'}(n)}{n}\bigg\lbrace\sum\limits_{i\in\mathcal{S}}\frac{N_{h'}(n,i)}{N_{h'}(n)}H\left(\frac{N_{h'}(n,i,\cdot)}{N_{h'}(n,i)}\right)
	-\sum\limits_{i\in\mathcal{S}}\mu_2(i)H(P_2(\cdot|i))\bigg\rbrace<-\epsilon,
	\frac{N_a(n)}{n}>\frac{\delta}{2K}(1-\epsilon')\text{ for all }a\in\mathcal{A}\bigg)\nonumber\\
	&+\sum\limits_{a=1}^{K}P\left(\frac{N_a(n)}{n}\leq\frac{\delta}{2K}(1-\epsilon')\right).\label{eq:exp_bound_t_4(n)/n_1}
\end{align}\endgroup
From the analysis using the Azuma-Hoeffding inequality for bounded difference submartingales presented earlier, we know that each term inside the summation in \eqref{eq:exp_bound_t_4(n)/n_1} is exponentially bounded. The first term in \eqref{eq:exp_bound_t_4(n)/n_1} may be written as
\begingroup\allowdisplaybreaks\begin{align}
	&P\bigg(\frac{N_{h'}(n)}{n}\bigg\lbrace\sum\limits_{i\in\mathcal{S}}\frac{N_{h'}(n,i)}{N_{h'}(n)}H\left(\frac{N_{h'}(n,i,\cdot)}{N_{h'}(n,i)}\right)
	-\sum\limits_{i\in\mathcal{S}}\mu_2(i)H(P_2(\cdot|i))\bigg\rbrace<-\epsilon,
	~\frac{N_a(n)}{n}>\frac{\delta}{2K}(1-\epsilon')\text{ for all }a\in\mathcal{A}\bigg)\nonumber\\
	&\leq P\bigg(\bigg\lbrace\sum\limits_{i\in\mathcal{S}}\frac{N_{h'}(n,i)}{N_{h'}(n)}H\left(\frac{N_{h'}(n,i,\cdot)}{N_{h'}(n,i)}\right)
	-\sum\limits_{i\in\mathcal{S}}\mu_2(i)H(P_2(\cdot|i))\bigg\rbrace<-{\epsilon},
	~\frac{N_a(n)}{n}>\frac{\delta}{2K}(1-\epsilon')\text{ for all }a\in\mathcal{A}\bigg).\label{eq:exp_bound_t_4(n)/n_2}
\end{align}\endgroup
From Lemma \ref{lemma:convergence_of_ML_estimates}, we have the following almost sure convergences as $n\to\infty$:
\begingroup\allowdisplaybreaks\begin{align}
	\frac{N_{h'}(n,i,j)}{N_{h'}(n,i)}&\to P_2(j|i),\text{ for all }i,j\in\mathcal{S},\nonumber\\
	\frac{N_{h'}(n,i)}{N_{h'}(n)}&\to \mu_2(i),\text{ for all }i\in\mathcal{S}.
\end{align}\endgroup
Using the above convergences and the continuity of the Shannon entropy functional $H(\cdot)$, we get that there exist constants $\delta_1=\delta_1(\epsilon)$ and $\delta_2=\delta_2(\epsilon)$ such that the probability in \eqref{eq:exp_bound_t_4(n)/n_2} may be upper bounded by the probability
\begingroup\allowdisplaybreaks\begin{align}
	P\bigg(\exists~i,j\in\mathcal{S}\text{ such that }
	\bigg|\frac{N_{h'}(n,i,j)}{N_{h'}(n,i)}-P_2(j|i)\bigg|>\delta_1,\bigg|\frac{N_{h'}(n,i)}{N_{h'}(n)}-\mu_2(i)\bigg|>\delta_2,
	\frac{N_a(n)}{n}>\frac{\delta}{2K}(1-\epsilon')\text{ for all }a\in\mathcal{A}\bigg).\label{eq:exp_bound_t_4(n)/n_3}
\end{align}\endgroup
Noting that $(N_{h'}(n,i,j)-N_{h'}(n,i)P_2(j|i))_{n\geq 0}$ and $(N_{h'}(n,i)-N_{h'}(n)\mu_2(j|i))_{n\geq 0}$ are martingale sequences for all $i,j\in\mathcal{S}$, we may then express \eqref{eq:exp_bound_t_4(n)/n_3} as a probability of deviation of martingale sequences from zero, which may be exponentially bounded by using results from \cite[Theorem 1.2A]{Victor1999}.

\item We now upper bound the term in \eqref{eq:exp_bound_t_2(n)/n}. Towards this, we first pick $\epsilon_1>0$ satisfying
\begin{equation}
	0<\epsilon_1\leq \frac{\epsilon}{1+2\sum\limits_{i\in\mathcal{S}}\mu_1(i)H(P_1(\cdot|i))}.\label{eq:epsilon_1}
\end{equation}
Then, the following almost sure convergences hold for all $i,j\in\mathcal{S}$:
\begingroup\allowdisplaybreaks\begin{align}
\frac{N_h(n)}{n}&\to \lambda_\delta^*,\nonumber\\
\frac{N_h(n,i,j)}{N_h(n)}&\to \mu_1(i)P_1(j|i).	
\end{align}\endgroup
Following the steps leading up to \eqref{eq:liminf_t_2(n)/n_final}, we note that for every choice of $\epsilon'>0$, there exists $M=M(\epsilon')$ such that \eqref{eq:liminf_t_2(n)/n_final} holds. We now choose $\epsilon'$ such that
\begingroup\allowdisplaybreaks\begin{align}
	\frac{T_{2}(n)}{n}
    	&\geq \frac{N_h(n)}{n}\bigg\lbrace\bigg[\sum\limits_{i\in\mathcal{S}}\sum\limits_{j\in\mathcal{S}}(\mu_1(i)P_1(j|i)+\epsilon')
    	\log\frac{\mu_1(i)P_1(j|i)+\epsilon'}{\mu_1(i)+\epsilon'|\mathcal{S}|}\bigg]-\epsilon'\bigg\rbrace\nonumber\\
    	&\geq \frac{N_h(n)}{n}\bigg(\sum\limits_{i\in\mathcal{S}}\mu_1(i)(-H(P_1(\cdot|i)))\bigg)-\epsilon_1
\end{align}\endgroup
holds for all sufficiently large values of $n$, where the last line above follows from the continuity of the term within braces as a function of $\epsilon'$. We then have
\begingroup\allowdisplaybreaks\begin{align}
	&P\left(\frac{T_2(n)}{n}-\frac{N_h(n)}{n}\sum\limits_{i\in\mathcal{S}}\mu_1(i)(-H(P_1(\cdot|i)))<-\epsilon\right)\nonumber\\
	&\leq P\bigg(\frac{T_2(n)}{n}-\frac{N_h(n)}{n}\sum\limits_{i\in\mathcal{S}}\mu_1(i)(-H(P_1(\cdot|i)))<-\epsilon,
	\bigg\vert\frac{N_h(n)}{n}-\lambda_\delta^*\bigg\vert\leq\epsilon_1,\nonumber\\
	&\hspace{8cm}
	\bigg\vert\frac{N_h(n,i,j)}{N_h(n)}-\mu_1(i)P_1(j|i)\bigg\vert\leq\epsilon'\text{ for all }i,j\in\mathcal{S}\bigg)\nonumber\\
	&+P\bigg(\bigg\vert\frac{N_h(n)}{n}-\lambda_\delta^*\bigg\vert>\epsilon_1\bigg)+\sum\limits_{i,j\in\mathcal{S}}P\bigg(\bigg\vert\frac{N_h(n,i,j)}{N_h(n)}-\mu_1(i)P_1(j|i)\bigg\vert>\epsilon'\bigg).\label{eq:exp_bound_t_2(n)/n_1}
\end{align}\endgroup
We now focus on the first term in \eqref{eq:exp_bound_t_2(n)/n_1}, and notice that for all sufficiently large values of $n$, this term may be upper bounded as
\begingroup\allowdisplaybreaks\begin{align}
	&P\bigg((\lambda_\delta^*+\epsilon_1)\sum\limits_{i\in\mathcal{S}}\mu_1(i)(-H(P_1(\cdot|i)))-\epsilon_1
	<-\epsilon+(\lambda_\delta^*-\epsilon_1)\sum\limits_{i\in\mathcal{S}}\mu_1(i)(-H(P_1(\cdot|i)))\bigg)\nonumber\\
	&\leq P\bigg(\epsilon_1>\frac{\epsilon}{1+2\sum\limits_{i\in\mathcal{S}}\mu_1(i)H(P_1(\cdot|i))}\bigg)\nonumber\\
	&=0,
\end{align}\endgroup
where the last line follows from the choice of $\epsilon_1$ in \eqref{eq:epsilon_1}. Exponential bounds for the remaining terms in \eqref{eq:exp_bound_t_2(n)/n_1} can be obtained similarly as in the analysis of the first term in \eqref{eq:exp_bound_t_4(n)/n_1}.

Lastly, for the terms in \eqref{eq:exp_bound_t_1/n}, \eqref{eq:exp_bound_t_3(n)/n},  \eqref{eq:exp_bound_t_5(n)/n} and \eqref{eq:exp_bound_t_6}, noting that the left-hand sides of the inequality inside the probability expression in all the three terms converge to zero a.s., similar procedures as used above for \eqref{eq:exp_bound_t_2(n)/n} and \eqref{eq:exp_bound_t_4(n)/n} may be used to obtain exponential upper bounds.
\end{enumerate}
This completes the proof of the lemma.
\end{IEEEproof}

Using the result of Lemma \ref{lemma:exp_bound} in \eqref{eq:uniform_integrability_1}, we get that there exist constants $\theta>0$ and $0<B<\infty$ independent of $L$ such that the following holds:
\begingroup\allowdisplaybreaks\begin{align}
\limsup\limits_{L\to\infty}E^\pi\bigg[\exp\bigg(\frac{\tau(\pi)}{\log L}\bigg)\bigg|C\bigg]		&{\leq} \exp\bigg(\frac{3}{\frac{\delta}{2K}(D(P_1||P_\delta|\mu_1)+D(P_2||P_\delta|\mu_2))}\bigg)
	+\limsup\limits_{L\to\infty}\sum\limits_{n\geq l(L,\delta)}B\exp\bigg(\frac{n+1}{\log L}-n \theta\bigg)\nonumber\\
	&<\infty,
\end{align}\endgroup
thus establishing that the family $\{\tau(\pi^\star(L,\delta))/\log L:L\geq 1\}$ is uniformly integrable.

Combining the above result on uniform integrability along with the asymptotic bound in \eqref{eq:almost_sure_upper_bound_for_policy_pi_star} yields the desired upper bound in \eqref{eq:upper_bound}, thus completing the proof of the proposition.
\end{IEEEproof}

\section{Summary}\label{sec:conclusions}
We analysed the asymptotic behaviour of policies for a problem of odd arm identification in a multi-armed rested bandit setting with Markov arms. The asymptotics is in the regime of vanishing probability of error. Our setting is one in which the \textcolor{black}{transition law of neither the odd arm nor the non-odd arms is known}. We derived an asymptotic lower bound on the expected stopping time of any policy as a function of error probability. We identified an explicit configuration-dependent constant in the lower bound. {Furthermore, we proposed a scheme that (a) is a modification of the classical GLRT, and (b) uses an idea of ``forced exploration'' from \cite{albert1961sequential}.} This scheme takes as inputs two parameters: $L\geq 1$ and $\delta\in(0,1)$. We showed that (a) for a suitable choice of $L$, the probability of error of our scheme can be controlled to any desired tolerance level, and (b) by tuning $\delta$, the performance of our scheme can be made arbitrarily close to that given by the lower bound for vanishingly small error probabilities. {In proving the above results, we highlighted how to overcome some of the key challenges that the Markov setting offers in the analysis. To the best of our knowledge, the odd arm identification problem (or variants like the best arm identification) in the Markov observations setting have not been analysed in the literature. Our analysis of the rested Markov setting is a key first step in understanding the different case of restless Markov setting, which is still open.}

\bibliographystyle{IEEEtran}
\bibliography{IEEEabrv,oai_rested_arms}

%








\end{document}